\documentclass[fleqn,10pt]{wlscirep}
\usepackage[utf8]{inputenc}
\usepackage[T1]{fontenc}
\usepackage{graphicx, caption, subcaption}
\usepackage{enumitem}
\usepackage{float}
\usepackage[normalem]{ulem}
\usepackage{multicol}
\usepackage{picture}

\title{Network coevolution drives segregation and enhances Pareto optimal equilibrium selection in coordination games}

\author[1,2,*]{Miguel A. Gonz\'alez Casado}
\author[1,3]{Angel S\'anchez}
\author[2]{Maxi San Miguel}

\affil[1]{Grupo Interdisciplinar de Sistemas Complejos (GISC), Universidad Carlos III de Madrid, 28911 Leganés, Spain}
\affil[2]{Institute for Cross-Disciplinary Physics and Complex Systems IFISC (CSIC-UIB), Campus Universitat Illes Balears, 07122 Palma de Mallorca, Spain}
\affil[3]{Instituto de Biocomputaci\'on y F\'\i sica de Sistemas Complejos (BIFI), Universidad de Zaragoza, 50018 Zaragoza, Spain}\affil[*]{Corresponding author; e-mail: miguelangel.gonzalezc@outlook.es}

\begin{abstract}
In this work we assess the role played by the dynamical adaptation of the interactions network, among agents playing Coordination Games, in reaching global coordination and in the equilibrium selection. Specifically, we analyze a coevolution model that couples the changes in agents' actions with the network dynamics, so that  while agents play the game, they are able to sever some of their current connections and connect with others. We focus on two action update rules: Replicator Dynamics (RD) and Unconditional Imitation (UI), and we define a coevolution rule in which, apart from action updates, with a certain \textit{rewiring probability p}, agents unsatisfied with their current connections are able to eliminate a link and connect with a randomly chosen neighbor. We call this probability to rewire links the `network plasticity'. We investigate a Pure Coordination Game (PCG), in which choices are equivalent, and on a General Coordination Game (GCG), for which there is a \textit{risk-dominant} action and a \textit{payoff-dominant} one. Changing the plasticity parameter, there is a transition from a regime in which the system fully coordinates on a single connected component to a regime in which the system fragments in two connected components, each one coordinated on a different action (either if both actions are equivalent or not). The nature of this fragmentation transition is different for different update rules. Second, we find that both for RD and UI in a GCG, there is a regime of intermediate values of plasticity, before the fragmentation transition, for which the system is able to fully coordinate on a single component network on the payoff-dominant action, i. e., coevolution enhances payoff-dominant equilibrium selection for both update rules.
\end{abstract}
\begin{document}
\flushbottom
\maketitle
\thispagestyle{empty}
\section*{Introduction}

Coordination Games, within  Evolutionary Game Theory\cite{SelectionRules}, is a well established theoretical framework to study social group formation\cite{Buskens2008}. In this paper we study two-choice Coordination Games\cite{SelectionRules} played in pairs by agents embedded in a network of contacts, and we will use $A$ and $B$ to refer to the two possible actions each agent can choose.  The characteristic feature of these games is that agents receive a larger payoff if they choose the same action among the two possible ones (either $A$ or $B$) than if they choose different actions. Therefore, agents receive a larger payoff for \textit{coordinating} on one of the actions. In this sense, the key point is selecting the same action than others, more than the action itself. If the payoff received for coordinating on $A$ or $B$ is equivalent we are in a special type of Coordination Game, a Pure Coordination Game (PCG). If, instead, both actions are not equivalent and the payoffs are different depending on the action they coordinate on, we talk about a General Coordination Game (GCG). In this case agents might receive a larger payoff coordinating on one action, but that might be riskier.  An example of this last case could be mass mobilizations\cite{Buskens2008}. We will delve into both types of games in the following section. In a GCG the problem of equilibrium selection arises\cite{UIpayoffdominant}${}^{,}$\cite{topology1}${}^{,}$\cite{topology2}${}^{,}$\cite{SimonCoordination}${}^{,}$\cite{PayoffTopology}: agents can coordinate either on the \textit{payoff-dominant} action or on the \textit{risk-dominant} action (one can find in the next section a deep explanation of the concepts of \textit{risk} and \textit{payoff-dominance}). The questions of when global coordination is reached depending on the structure of the network of interaction among agents\cite{PCGbestresponse} and which equilibrium is selected has been recently reviewed and studied in detail under different evolutionary rules in static complex networks\cite{MaxiPrincipal}.

Nonetheless, real-world interaction networks are dynamic and change over time. Thus, we need to go beyond previous studies in which the network of interactions is static\cite{MaxiPrincipal}. Specifically, if one agent is using one action and one of its neighbors is using the other one, the fact that they are connected may indeed influence the agent to rethink its choice. However, the agent may be also motivated to break this interaction to establish a more profitable one with another agent.  A general model of this situation should consider both the influence of the network over the game dynamics, and the influence of the game over the network dynamics. The concept of a dynamical framework of interactions was introduced by M.W. Macy \cite{Macy} proposing a \textit{new research agenda in which the structure of the network is no longer a given but a variable...to explore how a social structure might evolve in tandem with the collective action it makes possible}. This framework goes now under the name of \textit{coevolution models}, introduced in the context of Game Theory for the Prisoner´s Dilemma \cite{introductionofcoevolution,AJS,PRE} and reviewed in this context in Ref. \cite{Perc2010}. The purpose of this paper is to study the consequences of coevolution on coordination and equilibrium selection, building upon the results for static networks\cite{MaxiPrincipal}. Previous results\cite{Buskens2008,BuskensExtension,CoevolutionCoordination1,CoevolutionCoordination2,CoevolutionCoordination3,Pestelacci} on this problem suggest that the ability to coordinate, and to coordinate on one or another equilibrium, will be largely affected by the introduction of a dynamical modification of the network's structure. 

Our study is based on computational simulations in which we consider a population of agents playing the game and we track the action each of the agents is using at each moment. Agents have an action and are embedded within a network of interactions that restricts which individuals each agent is allowed to play with. Each pairwise interaction between two agents constitutes an individual game. Agents play the same game using their own action with all of their neighbors, accumulating an aggregated payoff as the result of the sum of individual payoffs obtained from each game. In this context, actions \textit{compete} and evolve in time, reproducing and taking the place of other actions among the agents within the population, and while this competition takes place, we let the network of interactions evolve in time. This competition between actions takes place microscopically in terms of \textit{update rules} following a random asynchronous update. The update rule determines how the agent changes action. Basically, update rules merge together the agents' perception of the surrounding environment, their expectations and beliefs, their possibility to make mistakes, etc. and translate them into action updates within the agents during the game, mimicking natural selection or human learning. The choice of different update rules can lead to completely different outcomes\cite{MaxiPrincipal}${}^{,}$\cite{addedAnxo}. This implies that the result of any analysis done in the context of Evolutionary Game Theory is critically defined not only by the payoff matrix of the game considered, but also by the update rule chosen. There are different update rules in the literature\cite{SelectionRules}, but here we focus on two of them: Replicator Dynamics and Unconditional Imitation (see Methods for further information about these rules).

There are different coevolution rules studied in the literature\cite{Perc2010}. In this work we focus on coevolution referring to the temporal evolution of the links between nodes with rules that evaluate agents' payoffs in order to take decisions about link rewiring. The basic idea behind this rule is that agents that are unsatisfied with their current interactions can break these connections and look for more beneficial interactions with other agents. The rule goes as follows (specific details in the Methods' section): we consider one of two agents connected by a link and using two different actions. Then, with a certain \textit{rewiring probability} $p$, one of the agents cuts this link and creates another one with an agent \textit{randomly chosen} from the network. Basically, agents may prefer to stick to their choice (e.g., if they have many neighbors like them) or search for new partners to try to improve their payoff. We stress that, differently to other approaches, we let new links to form randomly even if the outcome worsens the situation of the agent before link removal. The rationale behind this choice is that in real-world interactions it is not realistic to consider that we only make new connections with other people if these connections lead to a better situation than before. This is our intention beforehand, but we do not know \textit{a priori} how others will behave with us and the truth is that sometimes we are simply mistaken and we connect with people that worsen our situation. It should be mentioned that ``occasional mistakes when forming new links" does not imply that the rewiring is random. The reason for choosing random rewiring specifically is to explore the most extreme case of ignorance, the one that would oppose coordination most strongly. Other rewiring models, including selective rewiring or triangular relations in other networks could be the subject of further studies of this problem. The crucial parameter our coevolution model is $p$, called \textit{network plasticity}\cite{AJS,PRE,CoevolvingVoterModel}, since it measures the ratio of the time scales of network evolution to the time scale of changes of the actions of the agents. The larger $p$, the faster the creation/deletion of links with respect to the change of actions of the agents. We will see that this ratio of time scales will  determine if the system is able to coordinate on a single connected component of the networks or if, conversely, it fragments into multiple connected components coordinated each one on different actions.

\subsection*{Coordination Games without coevolution}
We present below the Payoff Matrices defining a Pure Coordination Game (equation \ref{pcg_matrix}) and a General Coordination Game (equation \ref{gcg_matrix})\cite{MaxiPrincipal}:
\vspace{-12mm}
\begin{multicols}{2}
  \begin{equation}\label{pcg_matrix}
    \def\arraystretch{1.3}
    \begin{array}{lccll}
    \multicolumn{1}{l|}{}           & \multicolumn{1}{c|}{{\bf A}} & \multicolumn{1}{c|}{{\bf B}} &  &  \\ \cline{1-3}
    \multicolumn{1}{c|}{{\bf A}} & \multicolumn{1}{c|}{1}          & \multicolumn{1}{c|}{0}          &  &  \\ \cline{1-3}
    \multicolumn{1}{c|}{{\bf B}} & \multicolumn{1}{c|}{0}          & \multicolumn{1}{c|}{1}          &  &  \\ \cline{1-3}
                                    & \multicolumn{1}{l}{}            & \multicolumn{1}{l}{}            &  & 
    \end{array}
    \end{equation}\break
  \begin{equation}\label{gcg_matrix}
    \def\arraystretch{1.3}
    \begin{array}{lccll}
    \multicolumn{1}{l|}{}           & \multicolumn{1}{c|}{{\bf A}} & \multicolumn{1}{c|}{{\bf B}} &  &  \\ \cline{1-3}
    \multicolumn{1}{c|}{{\bf A}} & \multicolumn{1}{c|}{1}          & \multicolumn{1}{c|}{S}          &  &  \\ \cline{1-3}
    \multicolumn{1}{c|}{{\bf B}} & \multicolumn{1}{c|}{T}          & \multicolumn{1}{c|}{0}          &  &  \\ \cline{1-3}
                                    & \multicolumn{1}{l}{}            & \multicolumn{1}{l}{}            &  & 
    \end{array}
    \end{equation}
\end{multicols}
\vspace{-5mm}
In equation \ref{gcg_matrix} we need to impose the constraints $T<1$ and $S<0$. It is direct to see that in a PCG, both actions $A$ and $B$ are completely equivalent, and coordination on one of the two actions yields the highest payoff, irrespective of the action chosen. On the other hand, in a GCG the two actions are not equivalent anymore. The characteristic feature of this second kind of coordination game is that it can have two types of equilibrium: a \textit{payoff-dominant} equilibrium and a \textit{risk-dominant} equilibrium. Let us consider an agent that always chooses $A$, and let us assume that the rest of the agents choose $A$ or $B$ with equal probability. If our agent plays repeatedly the game with random agents, half of the times it will earn $1$ and half of the times it will earn $S$. Hence, on average, its payoff will be $(1+S)/2$. If, instead, it always chooses action $B$, on average it will earn $T/2$. Bearing this in mind, if $(1+S)/2>T/2$ we would be in a situation in which choosing $A$ would be the \textit{less risky} choice (since, on average, we would earn a higher payoff with this choice). Conversely, if $(1+S)/2<T/2$, $B$ is the less risky choice. Therefore, depending on the range of the parameters $(S,T)$ we are, actions $A$ or $B$ will be the less risky choice. We call this action, i. e. the one ensuring the largest average payoff, \textit{risk-dominant} action, and full coordination of the system on this action is called \textit{risk-dominant} equilibrium. On the other hand, coordination on $A$ always leads to a higher absolute payoff than coordination on $B$. Thus, we call action $A$ the \textit{payoff-dominant} action, and full coordination on this action \textit{payoff-dominant} equilibrium. Notice that we could be in the situation in which action $A$ is both \textit{payoff-dominant} and \textit{risk-dominant}, or in a situation in which action $A$ is \textit{payoff-dominant} and action $B$ is \textit{risk-dominant}. This explanation is summarized in Fig. 6 of this reference\cite{MaxiPrincipal}. In this Figure, it is easy to see that, provided $T<1$ and $S<0$, the parameter space is divided by the line $T=S+1$. Above this line, when $T> S+1$, $B$ is \textit{risk-dominant}, and below it, when $T< S+1$, $A$ is \textit{risk-dominant}, while in both cases action $A$ is \textit{payoff-dominant}. We are going to call this theoretical line the \textit{risk-dominant} transition line. 

As our point of departure, we are going to use the results obtained in this previous work\cite{MaxiPrincipal}, in which they investigate coordination and equilibrium selection in Coordination Games without coevolution. A reproduction of the main results we rely on can be found in the Supplementary Material (Section S1).
Regarding PCG, they focus on dependence of the coordination rate $\alpha$ (proportion of agents using action $A$) with the mean degree of the network $\langle k \rangle$. They obtained that for large values of $\langle k \rangle$, update rules completely determine the ability of the system to reach full coordination. For the Replicator Dynamics case, we see that for $\langle k \rangle\geq 5$ we see that the system is able to perfectly reach full coordination on one of the two actions. For the Unconditional Imitation case, nonetheless, even for large $\langle k \rangle$ there are lots of realizations in which the system reaches a frozen configuration far from full coordination. Although as we increase $\langle k \rangle$ the number of realizations ending in full coordination increases, we always have some frozen configurations contrary to what we obtained for the RD case.
On the other hand, with respect to the case of GCG, they focus on \textit{which equilibrium is selected} (when there is coordination) as a function of $S$ and $T$, for both update rules. For a value of $\langle k \rangle$ for which the system is, in principle, able to reach full coordination in a PCG ($\langle k \rangle = 30$ for instance), they obtained that, in a GCG, when we use the RD update rule, the system always coordinates on the \textit{risk-dominant} action. On the other hand, for the UI update rule, there is an intermediate region in the parameter space (above the \textit{risk-dominant} transition line) in which the system coordinates on the \textit{payoff-dominant} action even if this action is not \textit{risk-dominant}.

\section*{Results}
The main research question we want to address in this paper is \textit{How does the image presented for PCG and GCG change when we introduce coevolution?} Specifically:
\begin{itemize}
    \item \textit{Does coevolution change the ability of the system to reach full coordination, or does it break the system into connected components each one coordinated on a different action?}
    \item \textit{Does coevolution change the equilibrium selected?}
\end{itemize} 
Obviously, the answer to these questions will depend on three variables we need to fix beforehand: the value of $\langle k \rangle$, the update rule chosen (RD or UI), and the type of game considered (PCG or GCG). In the main text we will focus on high values of $\langle k \rangle$ for both types of games, for which the system is, in principle, able to fully coordinate on a single connected component in the absence of coevolution (even if some realizations, for the UI rule in a PCG, end up in a frozen configuration). Specifically, we are going to focus on the value $\langle k \rangle=30$, and then we will discuss what is the role played by $\langle k \rangle$ within the results obtained. For the PCG we will study both update rules separately, and we will focus on the following question: \textit{Does coevolution fragment the system before it is able to reach full coordination?} Also, for the UI case, in which some specific realizations freeze in an uncoordinated configuration in the absence of coevolution, we could also wonder if coevolution helps unfreeze these specific realizations, helping the system to reach full coordination. We analyzed if this was the case, but the change in behavior was not significant (but for completeness, we have included this part of the assessment in the Supplementary Material (Section S2)). Finally, regarding GCG, we will answer the question \textit{Does coevolution change the equilibrium selected?}, and we will focus on the RD rule, since results for the UI rule are qualitatively similar, and we will include the UI case in the Supplementary Material. 

All in all, in this paper we will take a high value of $\langle k \rangle$, we will determine if coevolution leads to a fragmentation of the network for a PCG, and we will determine the nature of this fragmentation depending on the update rule. Then, for a GCG game, we will assess the role of coevolution in equilibrium selection. As it is explained in the Methods, the procedure to answer these questions is simple: we set our Erdös-Rényi network with $N=1000$ nodes and randomly distributed actions and we let it evolve following our two modified update rules (RD and UI with coevolution) until it reaches a frozen state. Notice that, contrary to what we had in the case without coevolution, now the system is not able to reach a frozen configuration in which agents using different actions are connected but unable to evolve. Because of the way we have defined coevolution, the system always ends up breaking this link and rewiring to another neighbor. Hence, there are only two possible outcomes: either the system fragments in pieces, with each one of these pieces coordinated on one action, or the system reaches full coordination on a single connected component.

In order to answer our questions, fixing a value of the rewiring probability $p$, we measure the following quantities averaging over $3000$ realizations: $\alpha$, the number of connected components of the system and the size of these connected components. This will allow us to analyze whether the system reaches full coordination as a whole or, on the other hand, if it fragments in pieces, and for the latter case we will be able to characterize how this fragmentation occurs. Finally, we will repeat this process for different values of $p$ from $0$ to $1$.

\subsection*{Pure Coordination Games with Coevolution}
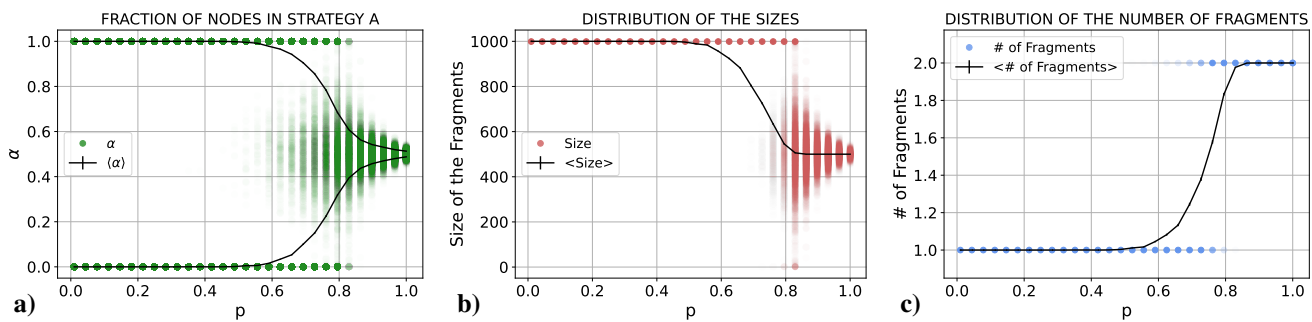
\begin{figure}[H]
    \centering
    \begin{minipage}[t]{0.33\linewidth}
    \begin{picture}(0.75\linewidth,0.75\linewidth)
        \put(0,0){\includegraphics[width=\linewidth]{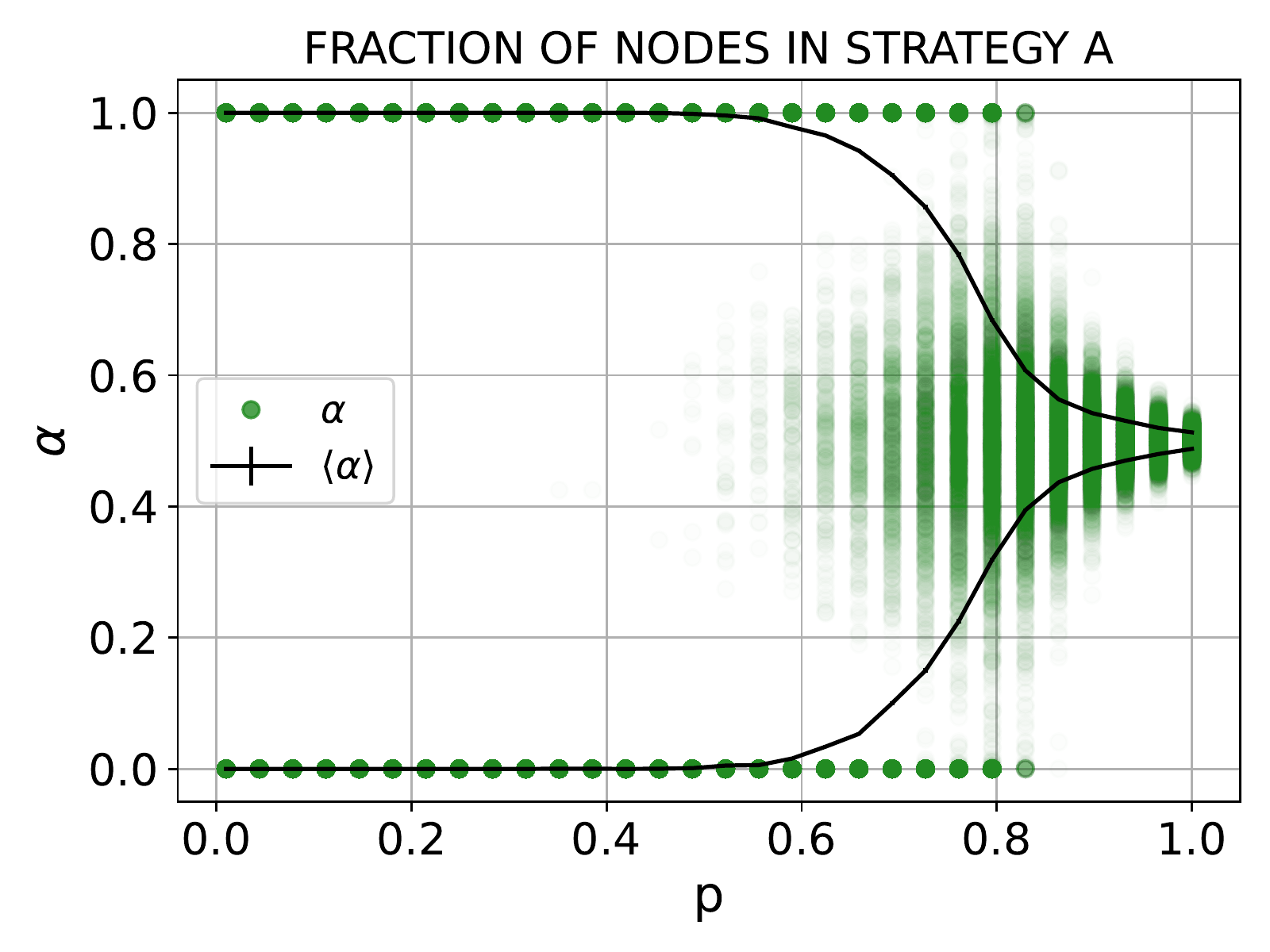}}
        \put(0.04\linewidth,0.04\linewidth){\textbf{a)}}
    \end{picture}
    \end{minipage}
    \begin{minipage}[t]{0.33\linewidth}
    \begin{picture}(0.75\linewidth,0.75\linewidth)
        \put(0,0){\includegraphics[width=\linewidth]{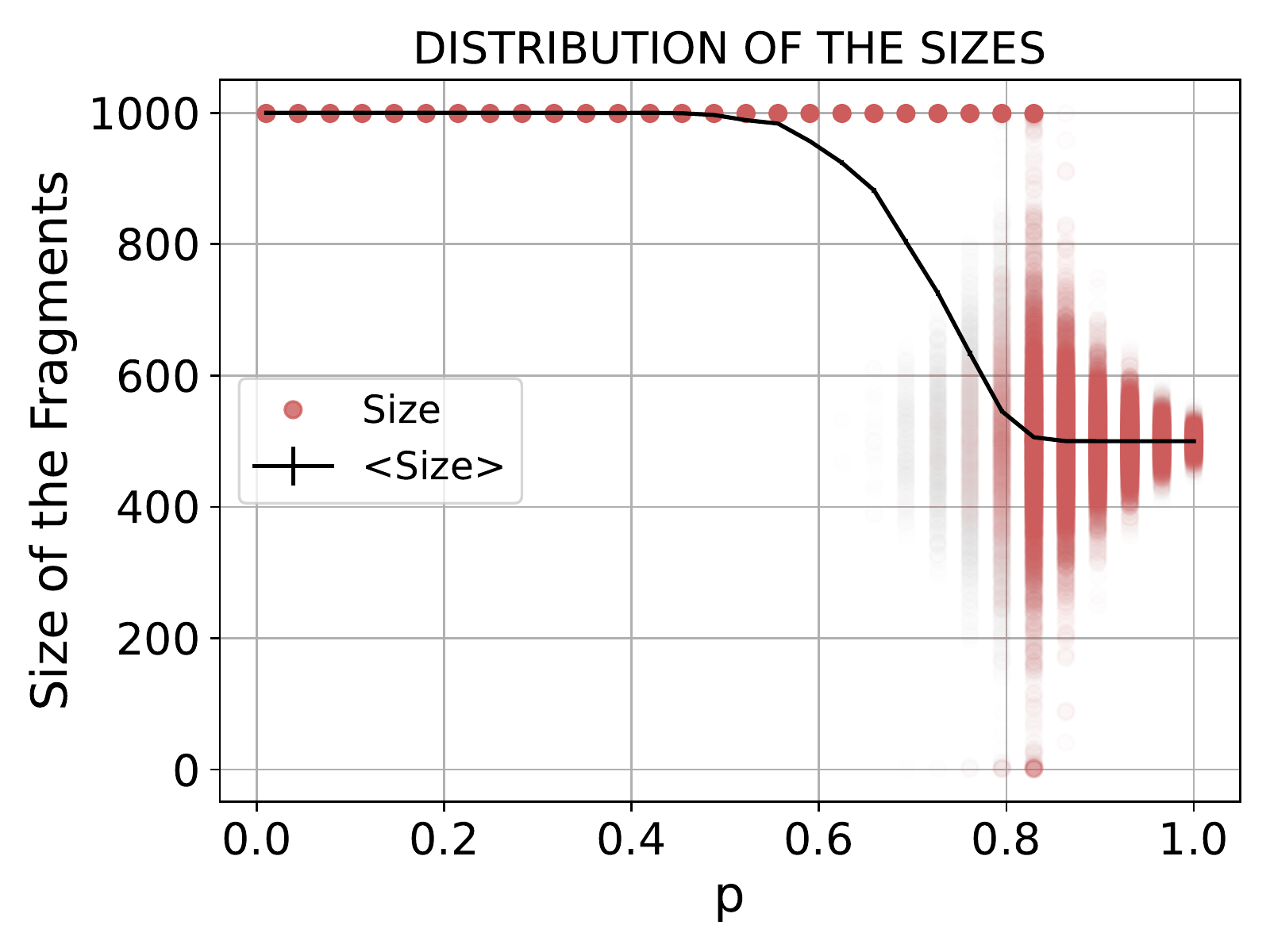}}
        \put(0.04\linewidth,0.04\linewidth){\textbf{b)}}
    \end{picture}
    \end{minipage}
    \begin{minipage}[t]{0.33\linewidth}
    \begin{picture}(0.75\linewidth,0.75\linewidth)
        \put(0,0){\includegraphics[width=\linewidth]{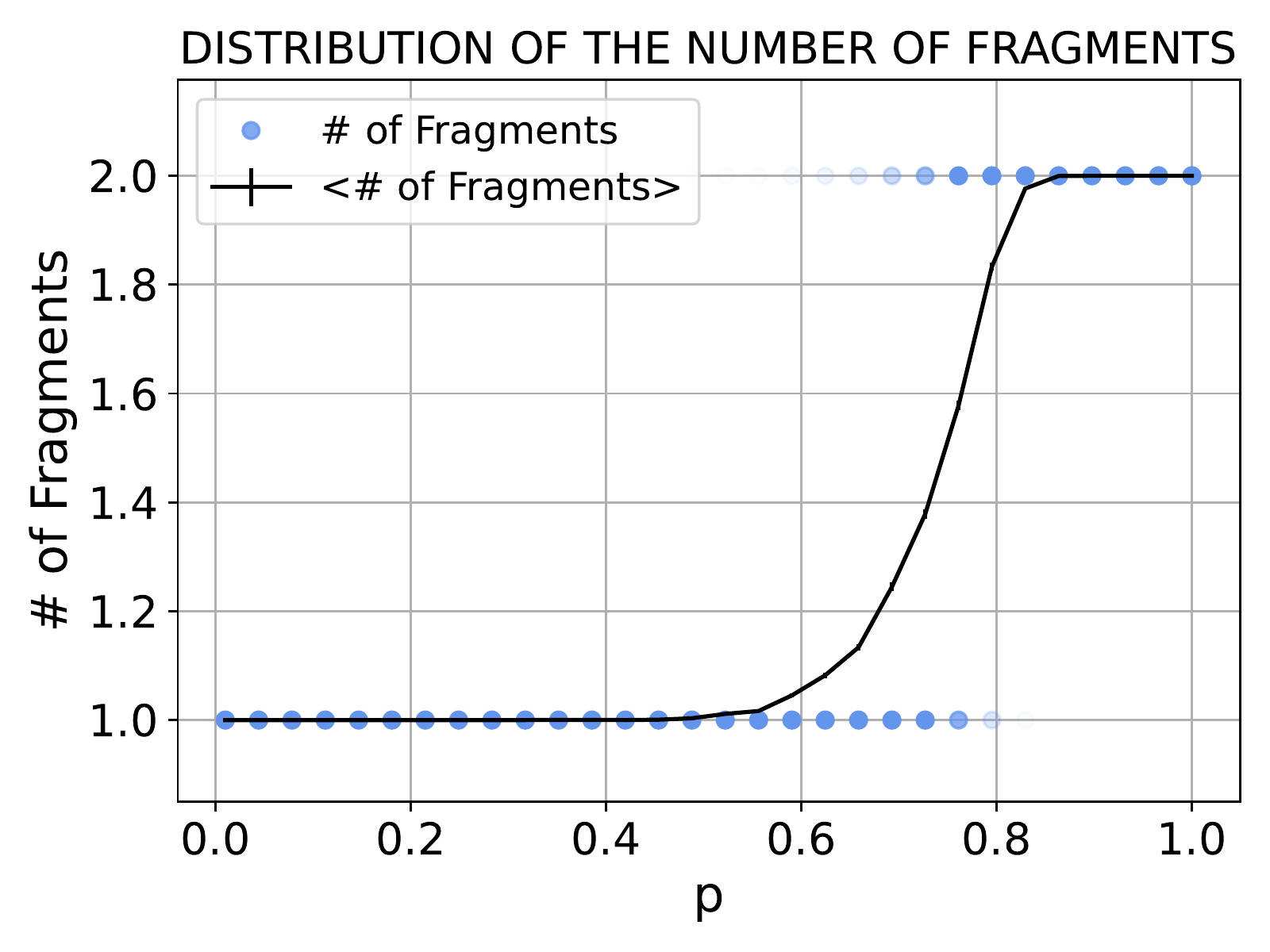}}
        \put(0.04\linewidth,0.04\linewidth){\textbf{c)}}
    \end{picture}
    \end{minipage}
    \caption{Results for a) $\alpha$, b) the size and c) the number of fragments (single realizations in green, red and blue, respectively, and average values in black) as a function of $p$ for a PCG played in an E-R network with $\langle k \rangle=30$ using the RD update rule.}
    \label{PCG_RDK30}
\end{figure}

\begin{figure}[H]
    \centering
    \begin{minipage}[t]{0.33\linewidth}
    \begin{picture}(0.75\linewidth,0.75\linewidth)
        \put(0,0){\includegraphics[width=\linewidth]{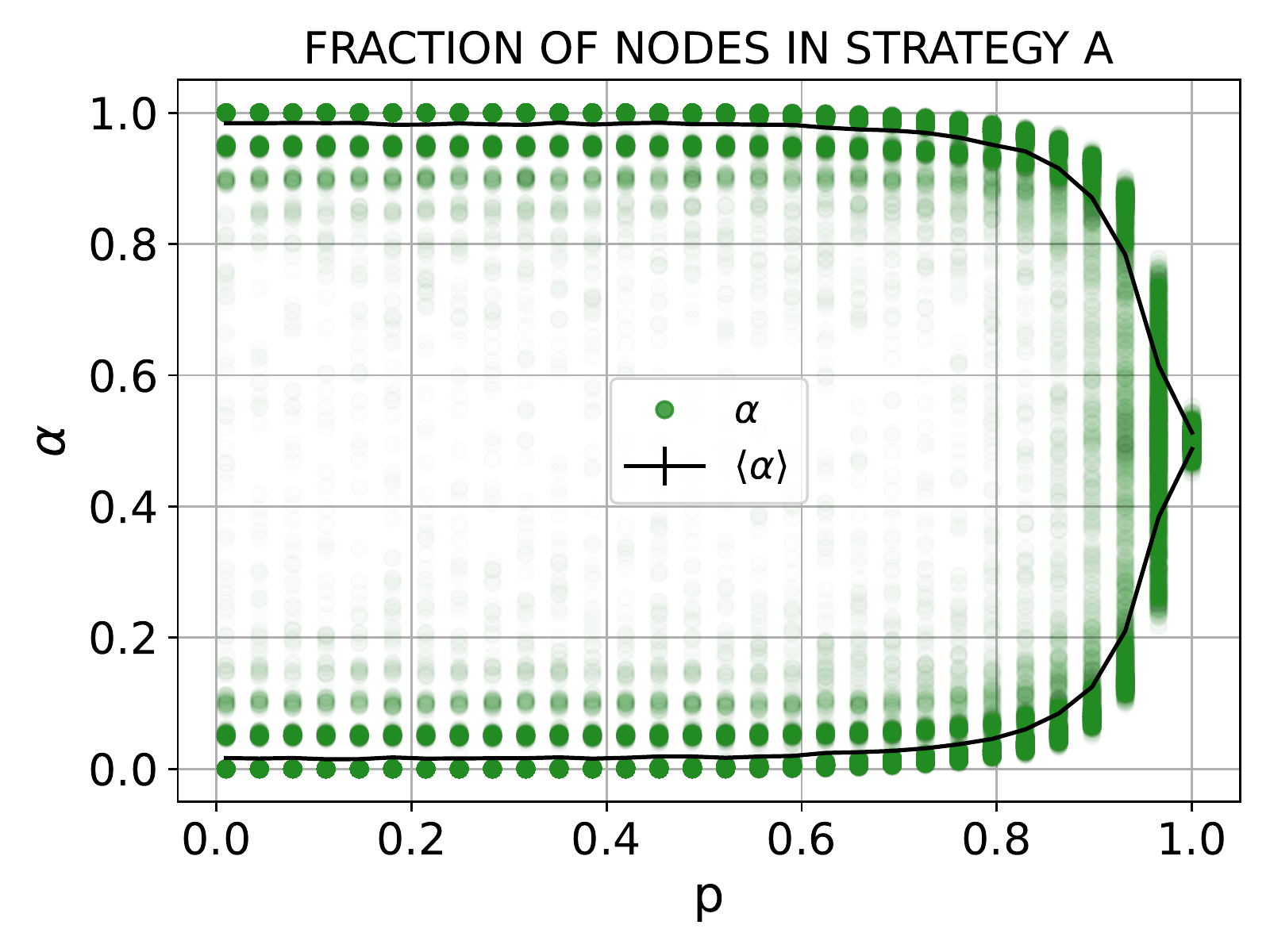}}
        \put(0.04\linewidth,0.04\linewidth){\textbf{a)}}
    \end{picture}
    \end{minipage}
    \begin{minipage}[t]{0.33\linewidth}
    \begin{picture}(0.75\linewidth,0.75\linewidth)
        \put(0,0){\includegraphics[width=\linewidth]{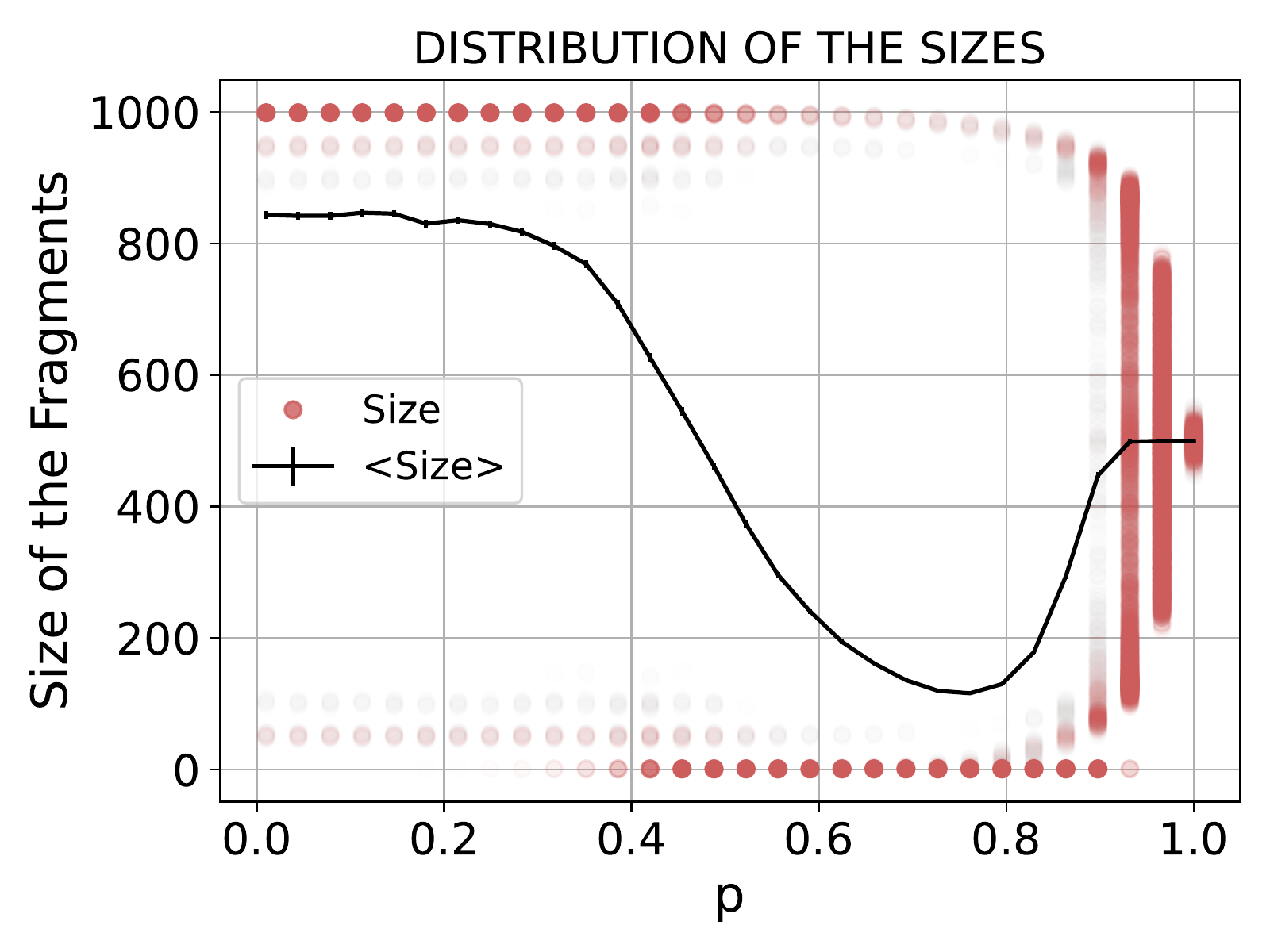}}
        \put(0.04\linewidth,0.04\linewidth){\textbf{b)}}
    \end{picture}
    \end{minipage}
    \begin{minipage}[t]{0.33\linewidth}
    \begin{picture}(0.75\linewidth,0.75\linewidth)
        \put(0,0){\includegraphics[width=\linewidth]{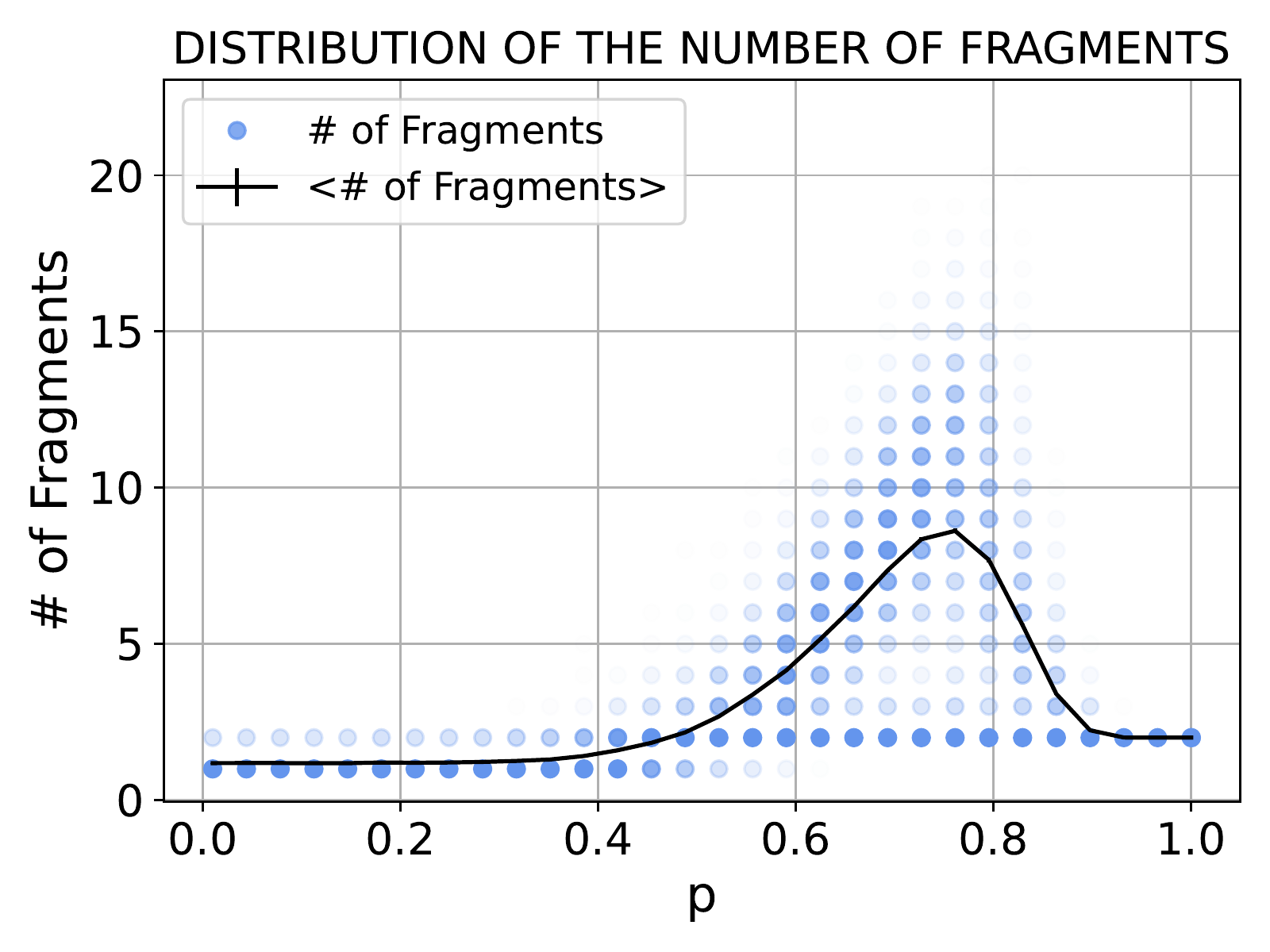}}
        \put(0.04\linewidth,0.04\linewidth){\textbf{c)}}
    \end{picture}
    \end{minipage}
    \caption{Results for a) $\alpha$, b) the size and c) the number of fragments (single realizations in green, red and blue, respectively, and average values in black) as a function of $p$ for a PCG played in an E-R network with $\langle k \rangle=30$ using the UI update rule.}
    \label{PCG_UIK30}
\end{figure}
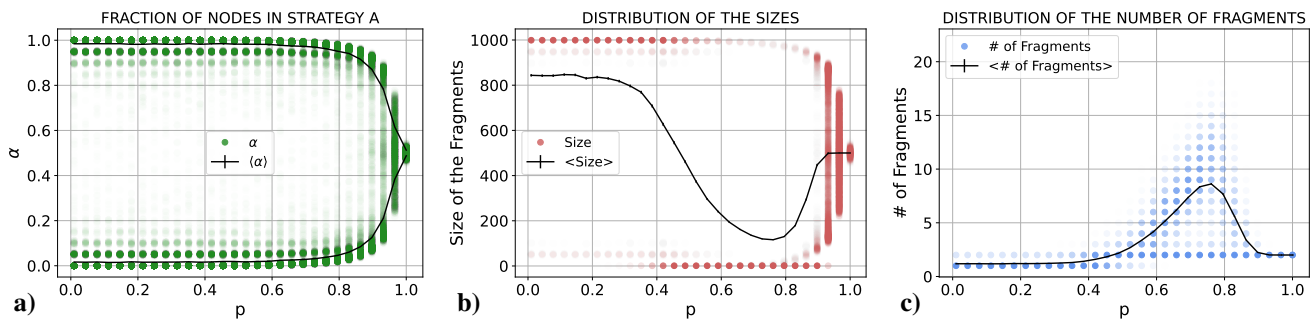

We present the results obtained for $\alpha$, the size of the fragments and the number of fragments as a function of $p$. In all cases we plot both the average value in black and the results of all the single realizations in colored points. These individual points are semi-transparent, allowing us to see in which parts of the graph points tend to concentrate more (the more intense the color, the larger the number of points concentrated there). The idea is to mimic the probability distribution function of the three quantities for each value of $p$ as if it were "seen from above". If the reader is interested in the actual probability distribution functions for these quantities, there is an example in the Supplementary Material (Section S3). Notice the large difference in the scales between Figs. \ref{PCG_RDK30}c and \ref{PCG_UIK30}c.

    \subsubsection*{Replicator Dynamics}
    
    Let us focus first on panels b) and c) of Fig. \ref{PCG_RDK30}. If we start from $p=0$ and increase $p$, the system still coordinates on a single connected component in every realization, as it did without coevolution (thus, coevolution does not play any significant role). Nonetheless, as $p$ approaches $0.7$ the systems sometimes still coordinates on a single connected component, and sometimes it starts to fragment into two connected components. At first these two connected components have a similar size, but as we continue increasing $p$ we cross a region in which the system is able to fragment into two pieces of virtually any size (sometimes a large connected component and a small one, sometimes of similar size, but always two fragments). Finally, when $p$ is close to one the system always fragments into two pieces of similar size and asymptotically for $p=1$ both components have the same size. If we now look at panel a) of Fig. \ref{PCG_RDK30}, we can see that when the system coordinates on a single connected component, half of the times it coordinates on one of the actions and the other half on the other action. On the other hand, when we are in the fragmentation regime each component coordinates on one of the two actions. All in all, we could say that there is a \textit{fragmentation transition} in $p$ from one component coordinated on a single action to two components coordinated each one on a different action. We will discuss later what is the nature of this transition, but before let us explore the UI case.

    \subsubsection*{Unconditional Imitation}
    
\begin{figure}[H]
        \centering
        \begin{minipage}[t]{0.33\linewidth}
                \includegraphics[width=\linewidth]{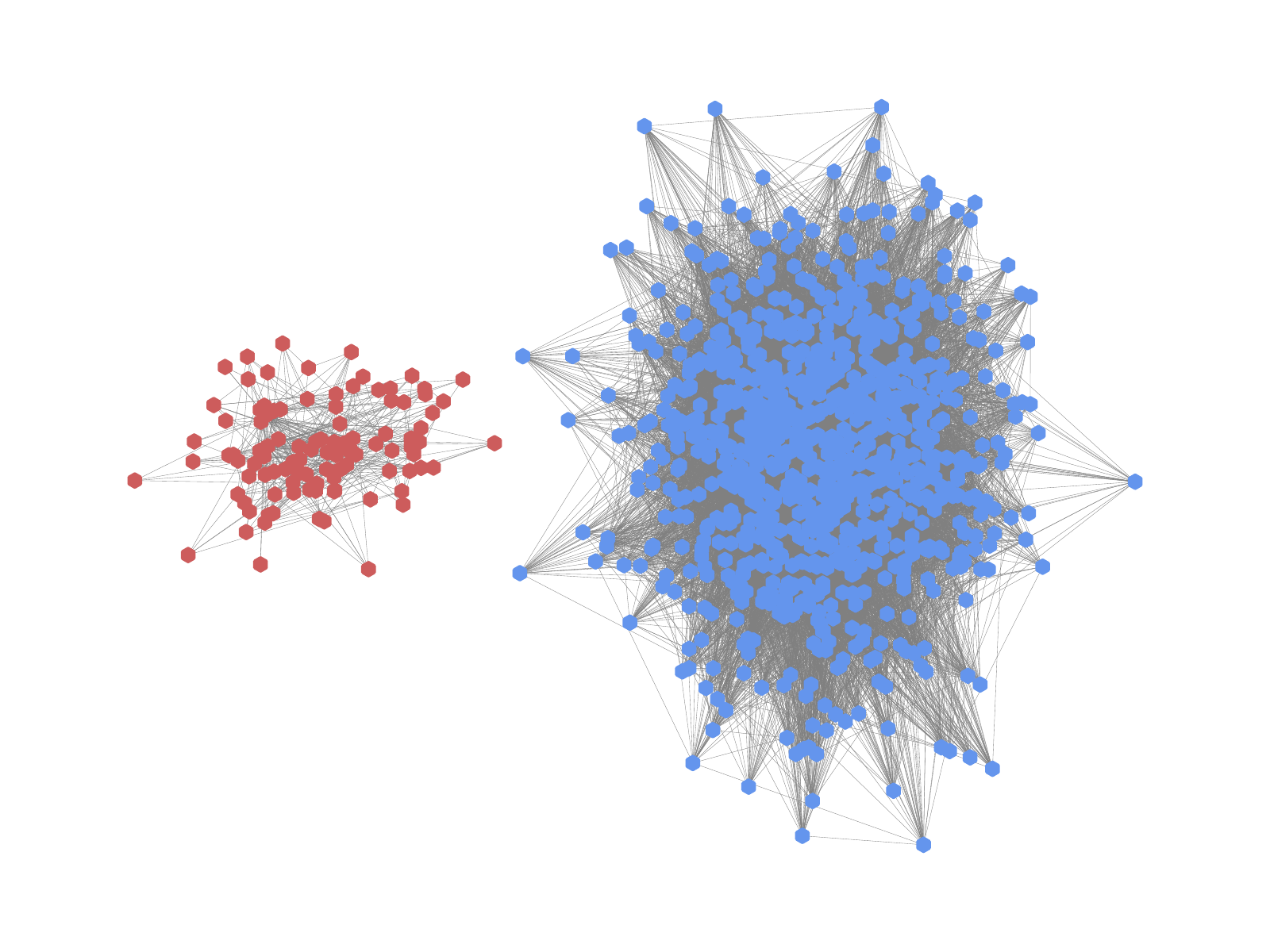}
            \end{minipage}
            \begin{minipage}[t]{0.33\linewidth}
                \includegraphics[width=\linewidth]{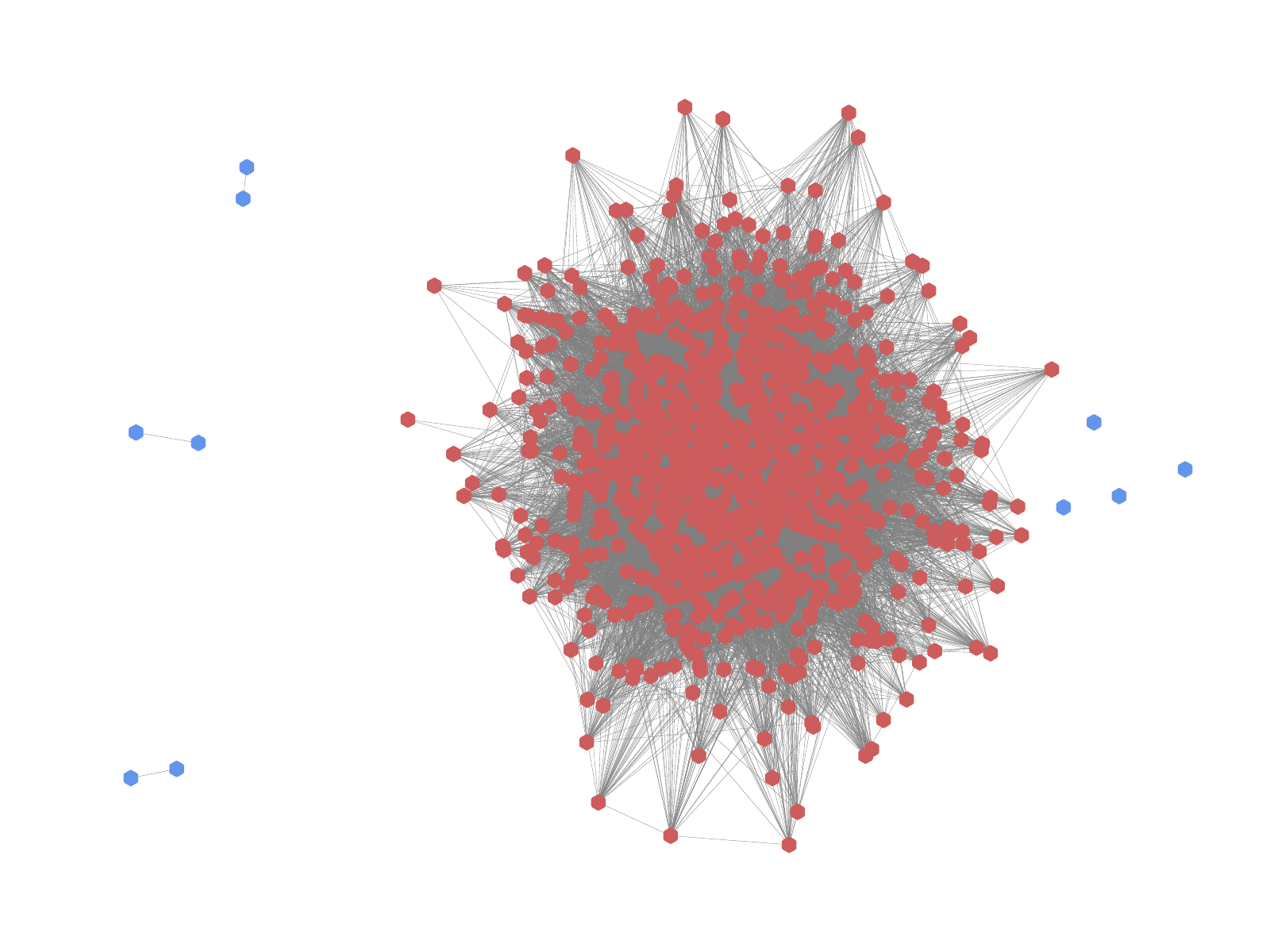}
            \end{minipage}
            \begin{minipage}[t]{0.33\linewidth}
                \includegraphics[width=\linewidth]{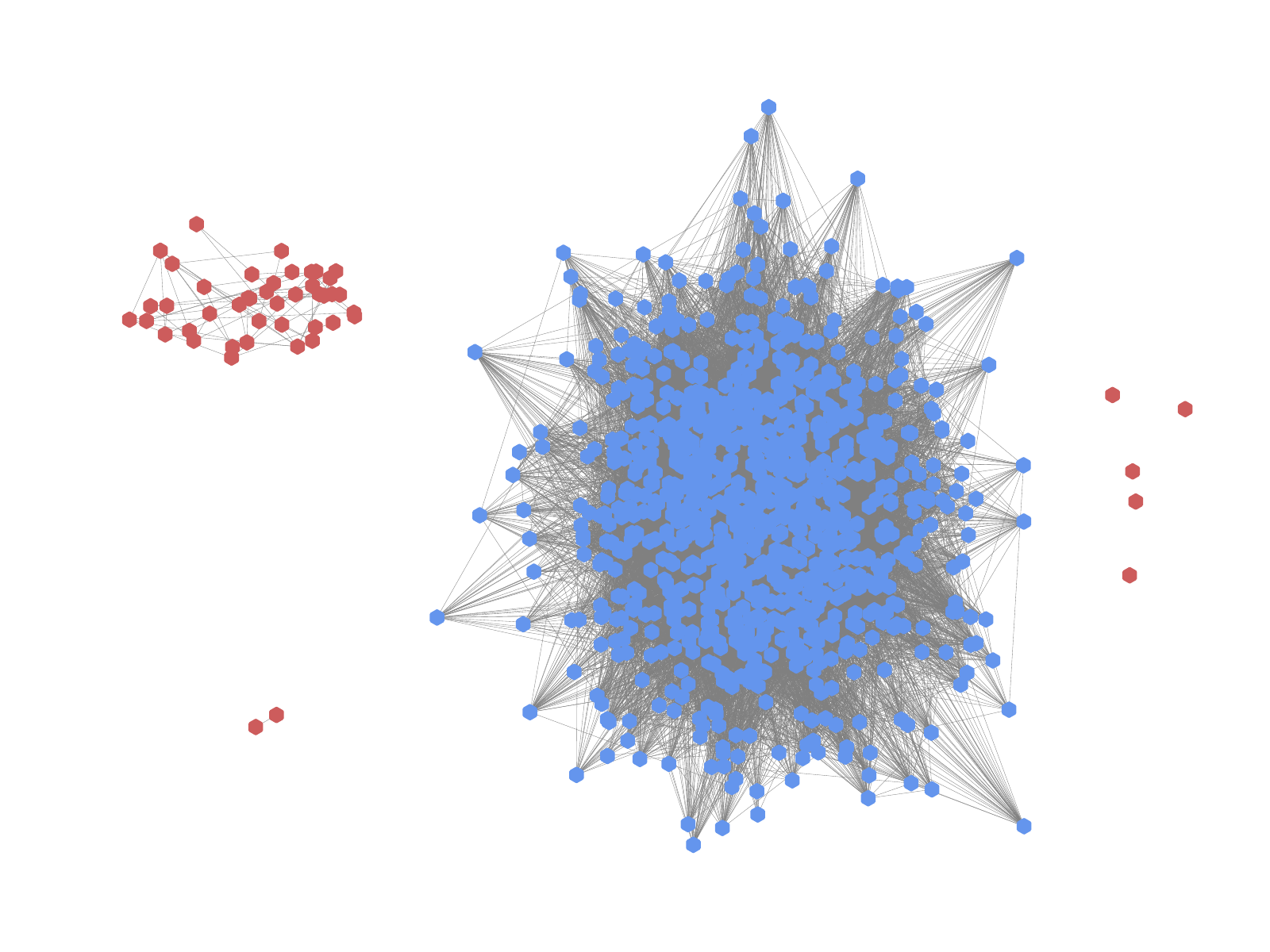}
            \end{minipage}
        \caption{Examples of a final state for $p=0.2$ (left), $p=0.7$ (center) and $p=0.85$ (right) using the UI rule.}
        \label{PCG_UIK30_networks}
    \end{figure}

    As in the RD case, we depict in Fig. \ref{PCG_UIK30} the results obtained for $\alpha$, the size of the fragments and the number of fragments as a function of $p$ for the UI case. It can be seen that the situation has drastically changed with respect to the RD case. In order to understand the forthcoming explanation, let us plot (Fig. \ref{PCG_UIK30_networks}) three final states obtained for three values of $p$ which may be difficult to understand otherwise. In this Figure, each color represents one action.
    If we start, as before, from $p=0$ and increase $p$, we see from panels b) and c) of Fig. \ref{PCG_UIK30} that for small values of $p$ the system fully coordinates in the majority of realizations. In the rest of realizations the system fragments into two connected components (an example of this case is plotted in the leftmost network of Fig. \ref{PCG_UIK30_networks}). The first striking fact about this behavior is that these two fragments always have almost the same size: the largest connected component around $950$ nodes and the smaller component around $50$ nodes. Notice that this is not the only possibility. With less probability they can have around $900$ and $100$ nodes, respectively. The important thing is that these values seem to be discrete. Bear in mind that we are in the case of low $p$, i. e., the final outcome is most likely due to competition between actions more than due to the coevolution. So, what we are really asking is: ``What mechanism of the UI rule is causing a fragmentation of the system in components of a very specific size? If a certain action survives, it means that, at least, a single agent had the largest payoff from all the neighbors of a second agent (so that this second agent has copied their action instead of copying the action of another neighbor), and if this has happened, on average 30 agents have been it the same situation with this ``successful" neighbor (since $\langle k \rangle=30$ and coevolution is almost negligible). Thus, we have that, if a certain action survives in a frozen state, at least in the most extreme case (where we have a central agent using an action and 30 neighbors connected to it using their action), this action is going to be used by, at least, 31 agents. However, those neighbors of the central, ``successful" agent are going to be connected between them and to other agents using the same action. Thus, we do not expect to have a component of specifically 30 nodes, but a component of a very specific size (in our case, we find $\sim$ 50 nodes).
        
    If we go to values of $p$ close to 1, as in the RD case, the system fragments into two pieces of similar size, one in each action. The main difference is in what happens between both extreme cases. Before, in RD, the transition from full coordination to fragmentation into two components happened through a mixed region in which sometimes the system fragmented into two pieces and sometimes the system reached full coordination. For UI, during the transition, although this fragmentation in two components still happens sometimes independently of $p$, other times the system fragments into multiple components. Specifically, going from low to high values of $p$ the system starts breaking into a giant component and multiple small pieces (an example of this case is shown in the central network of Fig. \ref{PCG_UIK30_networks}). If we continue increasing $p$, we reach a region in which we have a giant component, another component of not very small size and multiple small pieces (an example of this case is shown in the rightmost network of Fig. \ref{PCG_UIK30_networks}). Finally, as we reach $p=1$ these multiple small components disappear and the system breaks in only two pieces.
        
    \subsubsection*{Nature of the Fragmentation Transition}
\begin{figure}[H]
        \centering
        \begin{minipage}[t]{0.33\linewidth}
        \begin{picture}(0.75\linewidth,0.75\linewidth)
            \put(0,0){\includegraphics[width=\linewidth]{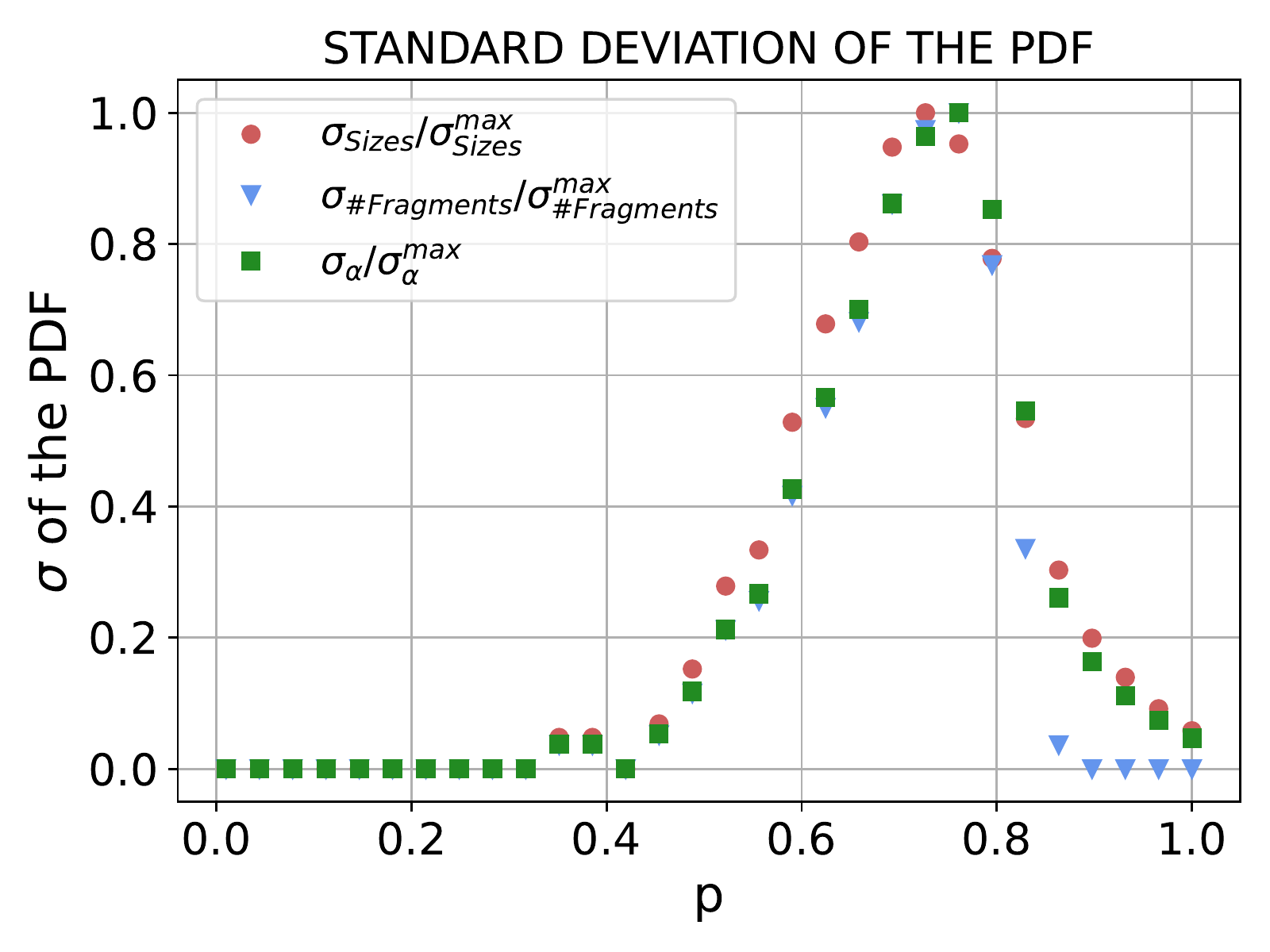}}
            \put(0.04\linewidth,0.04\linewidth){\textbf{a)}}
        \end{picture}
        \end{minipage}
        \begin{minipage}[t]{0.33\linewidth}
        \begin{picture}(0.75\linewidth,0.75\linewidth)
            \put(0,0){\includegraphics[width=\linewidth]{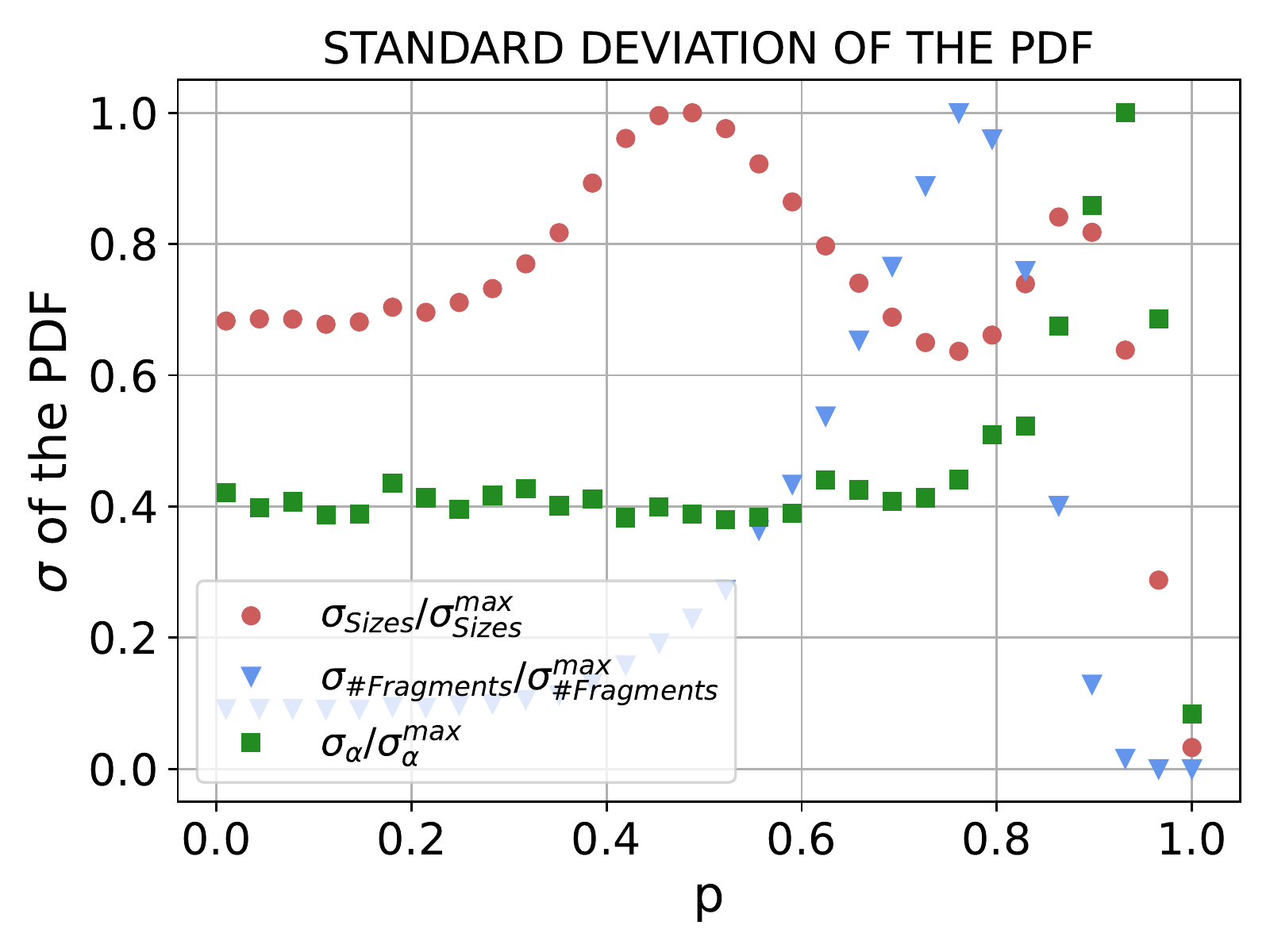}}
            \put(0.04\linewidth,0.04\linewidth){\textbf{b)}}
        \end{picture}
        \end{minipage}
        \caption{Variances of the distributions of $\alpha$, the size and the number of the fragments (divided by their maximum value) for each value of $p$ in the a) RD and b) UI cases.}
        \label{Sigma}
    \end{figure}

    We have seen that for both RD and UI cases there is a fragmentation transition in which the system goes from a regime in $p$ in which it is able to fully coordinate on a single connected component to a regime in $p$ in which it breaks into pieces. Nonetheless, it is clear how both transitions are essentially different. For the RD case the fragmentation is clean, from one component to two components without particularities. For the UI case, on the other hand, this transition happens through a crossover in which the system breaks into more than two fragments. In this section we analyze the variances of the distributions of $\alpha$, the size of the fragments and the number of fragments in order to gain insight into the nature of the transition. In Fig. \ref{Sigma} we plot both for RD and UI the variances of the three distributions divided by their maximum value respectively, in order to be able to compare the distributions between them.
    For the RD we have a clear result: both $\sigma_{\alpha}$ and $\sigma_{\# Fragments}$ have a single peak at $p = 0.76(3)$, and $\sigma_{Sizes}$ has it at $p = 0.73(3)$. These three peaks show that the system undergoes a clear transition from a regime in which the system coordinates on a single component to a regime in which the system fragments into two pieces, and this transition occurs at a certain $p$ within the range in which the three distributions have the maximum variance. It would be interesting to delve into the statistical nature of this transition, although it goes beyond the scope of this work. On the other hand, for the UI case the result is quite different. $\sigma_{Sizes}$ has a first peak around $p \sim 0.5$. Subsequently, it has a minimum coincident with a maximum in $\sigma_{\# Fragments}$ around $p \sim 0.79$, and then it rises again to have a second maximum almost coincident with a maximum in $\sigma_\alpha$ around $p \sim 0.9$. If we recall the interpretation of the previous results, the leftmost peak corresponds to the moment in which the system is not able anymore to coordinate on a single connected component, and most realizations end up in a large coordinated connected component and very small (1 - 2 nodes) fragments. At first, the number of small fragments is low, but as we increase $p$ we start having more and more fragments of this same size. This situation corresponds to the moment in which there is a peak in $\sigma_{\# Fragments}$ and a minimum in $\sigma_{Sizes}$ (it is the moment in which depending on the realization we could have from 2 up to 20 fragments, always a large connected component and groups of 1 - 2 nodes). Then, if we keep increasing $p$ we reach the second peak in $\sigma_{Sizes}$ coincident with the peak in $\sigma_\alpha$ at $p \sim 0.9$. In this regime, a second larger connected component is able to form, and as we increase $p$ this second large connected component is able to reach a size comparable to the size of the largest connected component and small fragments disappear. All in all, the variances in this second case tell us that there is not a sharp transition between the "single coordinated component" regime and the "two connected components" regime, but rather a crossover from one to the other.
    \subsubsection*{The role of <k>}
    
    In order to complete the analysis, we made an analogous assessment for other values of $\langle k \rangle$ within the range in which the system is able to reach full coordination in the absence of coevolution (specifically, $\langle k \rangle=10,~20$) to clarify the role $\langle k \rangle$ plays in the fragmentation transition. These results can be found in the Supplementary Material (Section S6).
    
    Basically, we obtained that $\langle k \rangle$ only changes the value of $p$ for which the transition occurs, but it does not qualitatively change the characteristics of the fragmentation transition. The smaller $\langle k \rangle$, the smaller the value of $p$ at which the transition takes place. This result is reasonable if we assess the transition from a microscopic point of view: a single connected component breaks into two pieces if all the links between both pieces are cut. The larger $\langle k \rangle$, the larger the number of links you need to cut to fragment the system. Hence, you need a larger probability of cutting a link for the fragmentation transition to happen. By and large, these results ensure that there is a wide range of values of $\langle k \rangle$ for which the conclusions drawn in the case $\langle k \rangle=30$ apply. The upper limit is set by the value of $\langle k \rangle$ for which no matter how high $p$ is the system never fragments. The lower limit is set by the value of $\langle k \rangle$ for which there is no possibility of reaching full coordination in the absence of coevolution, since ER networks are never connected in a single connected component for these values of $\langle k \rangle$. The study of this case goes beyond the scope of this work. 
    
    As a final remark, it is important to notice that these conclusions depend on the size of the system. We obtain these results for a system of $N=1000$ nodes. Thus, these values of $\langle k \rangle$ always need to be compared with the system size. We need to take this into account for two reasons: the first one is that for different system sizes we may obtain the same results but for different values of $\langle k \rangle$; and the second and more important reason is that some of our conclusions may be affected by finite-size effects and they may not apply in the thermodynamic limit for a large system size.

\subsection*{General Coordination Games with Coevolution}
As we mentioned above, we will focus only in the RD rule in order to simplify the analysis, and, at the end of the section, we present a brief comparison with the UI case. To begin with, let us recall what happens in the system for this rule in the absence of coevolution. This result can be found in the Supplementary Material. The main conclusion was that in the absence of coevolution, for each choice of $(S,T)$ the system coordinated always on the \textit{risk-dominant} action. Now, we add coevolution, and we want to answer basically three questions: 
\begin{itemize}
    \item \textit{Is there a fragmentation transition like in the case of a Pure Coordination Game?}
    \item \textit{If there is this fragmentation transition, does any of the fragments coordinate on the \textit{payoff-dominant} action, or all coordinate on the \textit{risk-dominant} action?}
    \item \textit{Can coevolution enable the whole system to coordinate on the payoff-dominant action?}
\end{itemize} 
Notice that, once we fix $\langle k \rangle=30$, we still have three free parameters: $p$, $S$, and $T$. To answer our questions, we take four specific pairs $(S,T)$ and we study the behavior of the system for each one when we vary $p$. Then, once we have understood the role of $p$, we will select a group of values of $p$ and study the variation of $S$ and $T$. 

    \subsubsection*{Fragmentation Transition}
    First, we are going to select our example pairs $(S,T)$. We will choose them in four representative regions of the parameter space of the system:
    \begin{itemize}
        \item $(S,T)=(-1.5,-1.5)$ The system coordinates on $A$ (which is both \textit{risk-dominant} and \textit{payoff-dominant}) in the absence of coevolution. This point is far from the transition line. 
        \item $(S,T)=(-2,-1)$ Point right at the transition line (where $A$ is \textit{payoff-dominant}, and there is no \textit{risk-dominant} action since both actions yield the same risk).
        \item $(S,T)=(-2.5,-0.5)$ The system coordinates on $B$ (which is the \textit{risk-dominant} action, while $A$ is the \textit{payoff-dominant} one) in the absence of coevolution. This point is far from the transition line. 
        \item $(S,T)=(-1.35,0)$ The system coordinates on $B$ (which is the \textit{risk-dominant} action, while $A$ is the \textit{payoff-dominant} one) in the absence of coevolution. This point is near the transition line.
    \end{itemize}
    As we will see, these four values are representative of the four types of behavior the system displays when we add coevolution. For the first three choices of $(S,T)$ we present below the measurements for $\alpha$ as a function of $p$. The fourth case will be analyzed in detail later.
    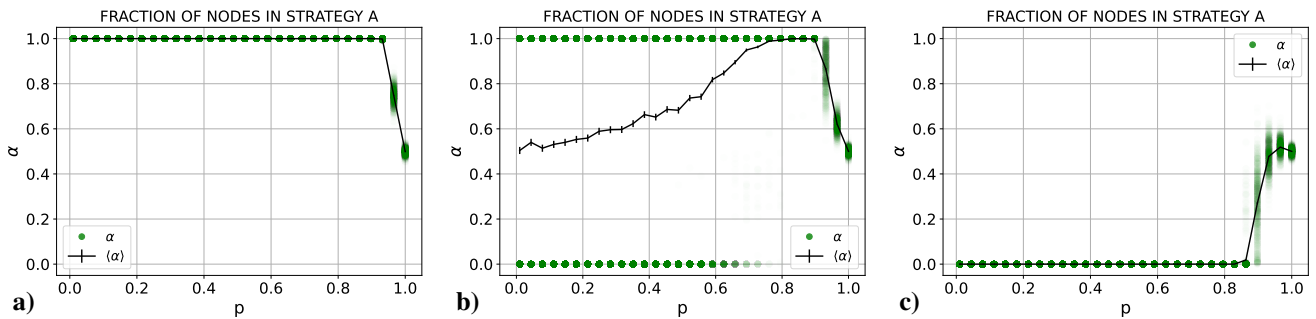
\begin{figure}[H]
        \centering
        \begin{minipage}[t]{0.33\linewidth}
        \begin{picture}(0.75\linewidth,0.75\linewidth)
            \put(0,0){\includegraphics[width=\linewidth]{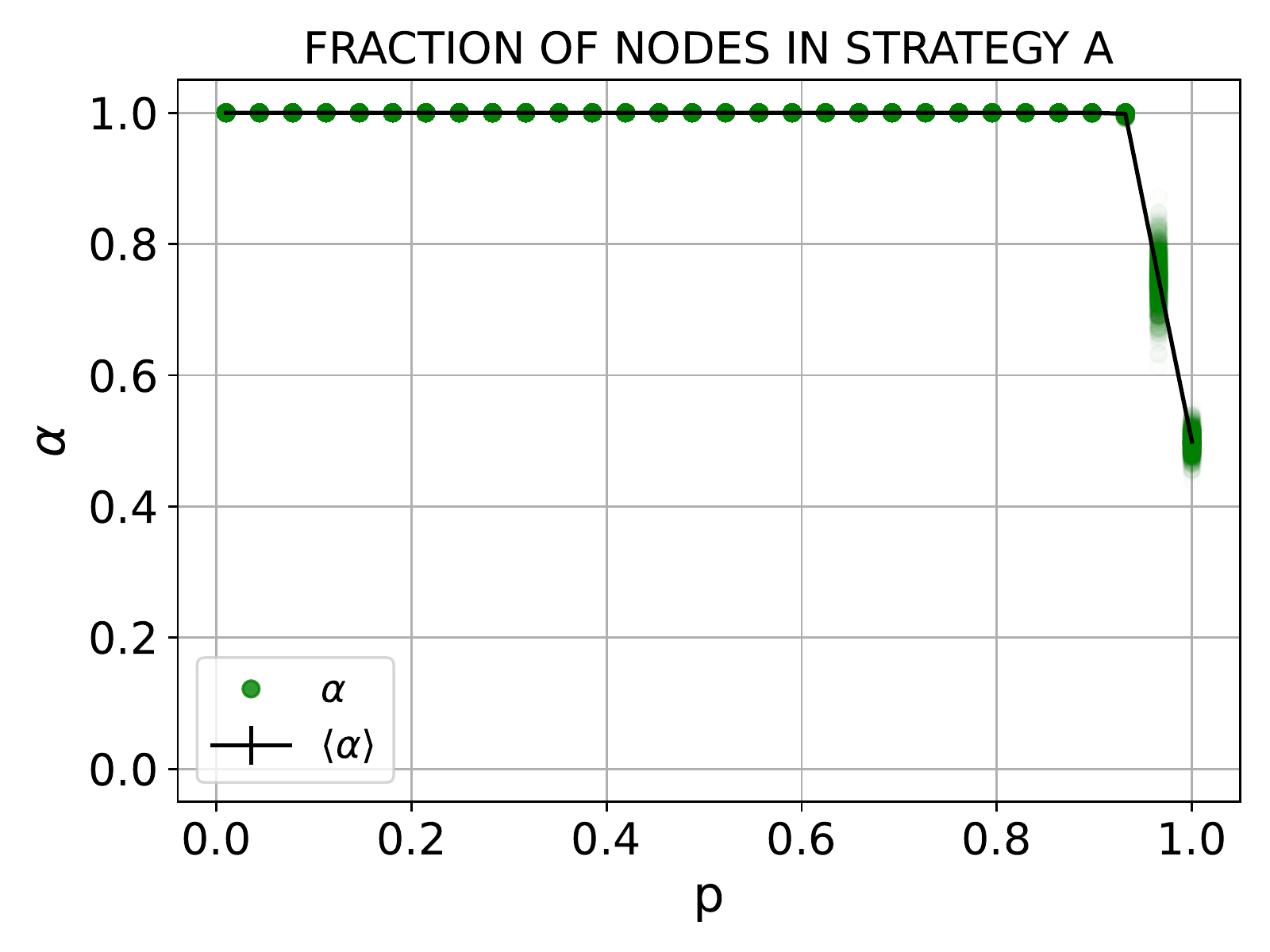}}
            \put(0.04\linewidth,0.04\linewidth){\textbf{a)}}
        \end{picture}
        \end{minipage}
        \begin{minipage}[t]{0.33\linewidth}
        \begin{picture}(0.75\linewidth,0.75\linewidth)
            \put(0,0){\includegraphics[width=\linewidth]{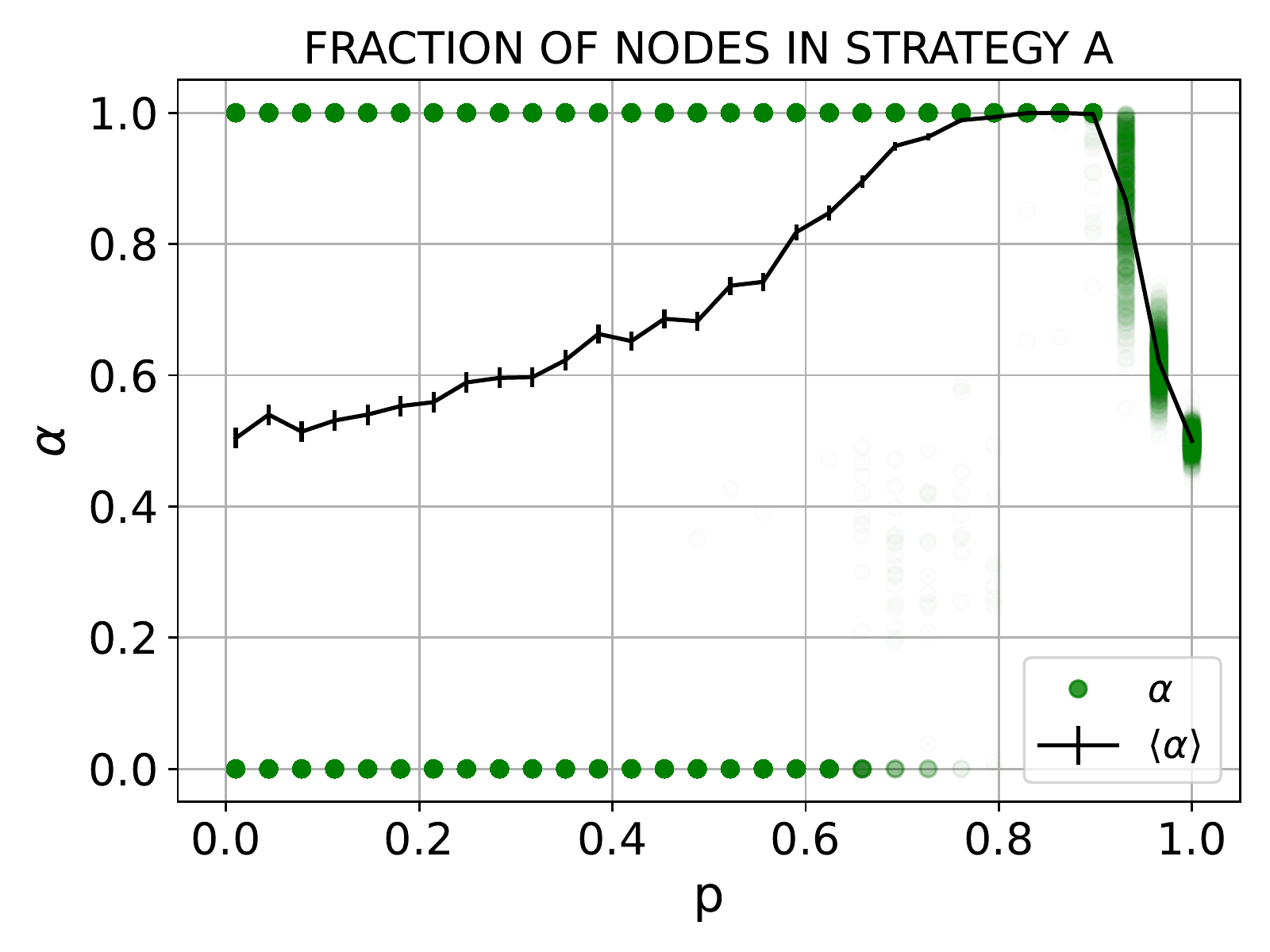}}
            \put(0.04\linewidth,0.04\linewidth){\textbf{b)}}
        \end{picture}
        \end{minipage}
        \begin{minipage}[t]{0.33\linewidth}
        \begin{picture}(0.75\linewidth,0.75\linewidth)
            \put(0,0){\includegraphics[width=\linewidth]{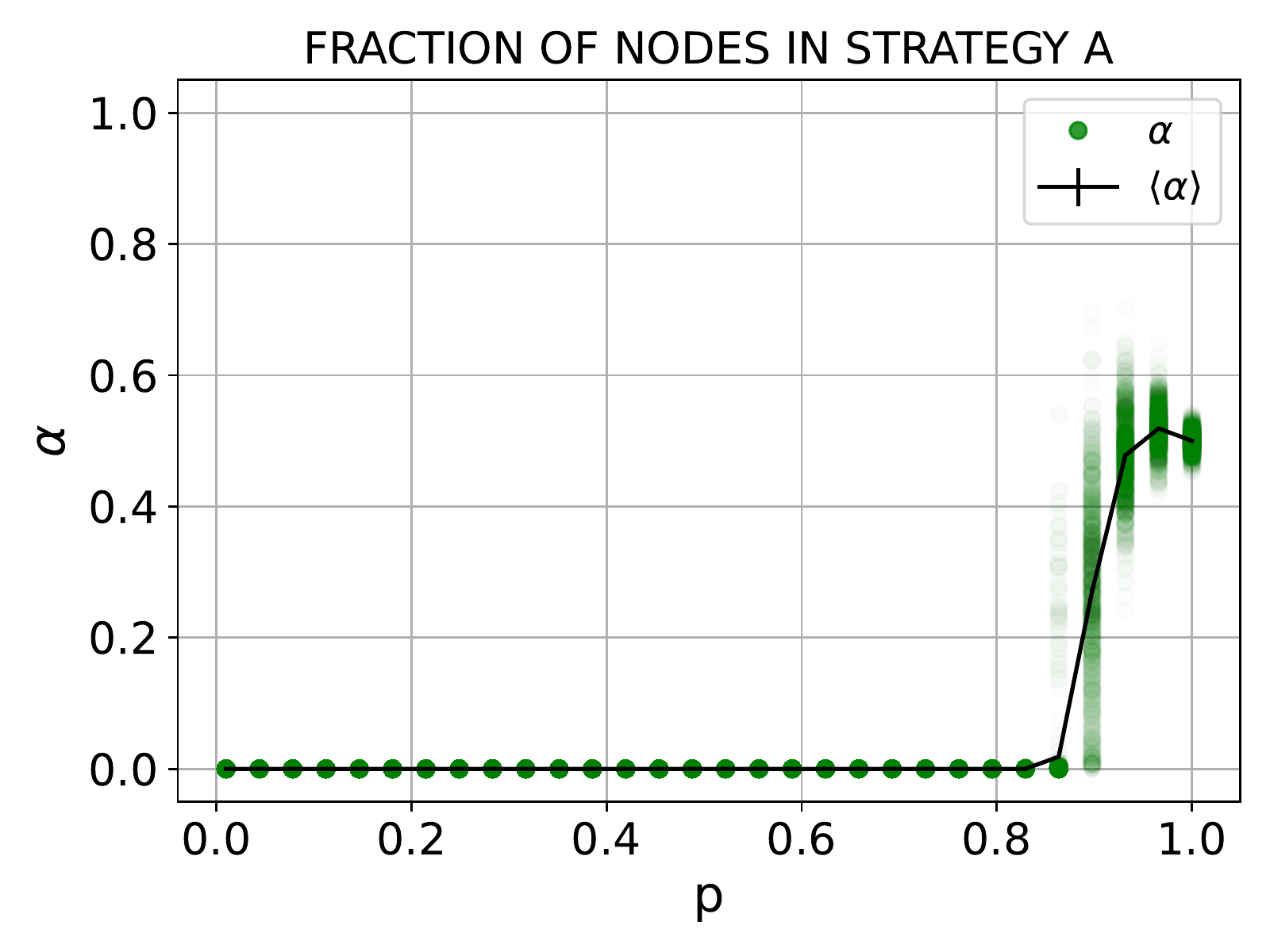}}
            \put(0.04\linewidth,0.04\linewidth){\textbf{c)}}
        \end{picture}
        \end{minipage}
        \caption{Results for $\alpha$ as a function of $p$ for a GCG with coevolution played in an E-R network with $\langle k \rangle=30$, using the RD update rule, whose parameters are a)  $(S,T)=(-1.5,-1.5)$, b) $(S,T)=(-2,-1)$ and c) $(S,T)=(-2.5,-0.5)$.}
        \label{GCG_C_fragmentation}
    \end{figure}
    It is important to mention that we have decided move the Figures of the size and the number of fragments as a function of $p$ to the Supplementary Material (Section S4). Basically, these Figures allow us to asses the nature of the fragmentation transition. What we have obtained is that for both rules there is a fragmentation transition, and this transition has the same features than the one observed for PCG. Thus, these two variables display the same qualitative behavior as well (panels b) and c) of Fig. \ref{PCG_RDK30}) and therefore the decision to move them to the SM to avoid redundancies. What is different for GCG is the asymmetry in the equilibrium selection, and this asymmetry can be analyzed directly from $\alpha$.
    
    If we look at the result for $\alpha$, specifically panels a) and c) of Fig. \ref{GCG_C_fragmentation}, for low and medium values of $p$ the system coordinates on the \textit{risk-dominant} action in a single component. Then, for large values of $p$, we observe the mentioned fragmentation transition in two components similar to the one observed for PCG (this can be confirmed looking at Figs. S6 and S7 of the Supplementary Material). For the fragmentation transition, we see that the big component coordinates on the \textit{risk-dominant} action, but the smaller component coordinates on the other action. In panel a) of Fig. \ref{GCG_C_fragmentation}, we are in the region of the parameters space in which action $A$ is both \textit{payoff-dominant} and \textit{risk-dominant}, so this result implies that there is a fraction of the population coordinating on the action which is neither \textit{payoff-dominant} nor \textit{risk-dominant}. In panel c) of this same Figure, on the other hand, we are in the region in which action $B$ is the \textit{risk-dominant} and action $A$ is the \textit{payoff-dominant}. Hence, coevolution is allowing a part of the system to coordinate on the \textit{payoff-dominant} action. These results answer our first two questions: there is a fragmentation transition of a similar nature to the one observed in PCG for the RD case, and this fragmentation transition enables one of the fragments to coordinate on the \textit{payoff-dominant} action even if it is not the \textit{risk-dominant} action.
    
    There is an even more interesting conclusion that can be drawn from these results. Notice that the choice of $(S,T)$ for panels a) and c) is symmetric with respect to the \textit{risk-dominant} transition line of the result without coevolution. Thus, it would be reasonable to expect a symmetric result in $\alpha$ with respect to $\alpha=0.5$. This is, we expect $\alpha$ in panel a) to be the mirror image of $\alpha$ in panel c) with respect to $\alpha=0.5$. The reason behind is that we are measuring at two points located at the same distance perpendicularly with respect to the \textit{risk-dominant} transition line. Hence, For low values of $p$ we would expect the same behavior observed without coevolution (coordination on the \textit{risk-dominant} action), and then a fragmentation transition. Nonetheless, if we compare results for both cases we see that the transition from $\alpha=1$ or $\alpha=0$ to $\alpha=0.5$ does not behave in a similar fashion. Furthermore, if we analyze panel b) of Fig. \ref{GCG_C_fragmentation}, with $(S,T)$ right at the transition line, we expect to observe a more or less symmetric result, but the result is balanced towards action $A$. Just by looking at these results it seems like the transition line has been displaced. If we had the transition line displaced towards the region in which action $B$ is the \textit{risk-dominant} action, that would explain the differences between panels a) and c) (now the first pair $(S,T)$ would be further from the transition line than the third pair $(S,T)$) and that would also explain the unbalance in panel b) (now this second pair $(S,T)$ is close to the transition line but still in the region in which $A$ is the \textit{risk-dominant} action). 
    
    In order to avoid confusions, let us stress that in the Figure included in the Supplementary Material (RD for a GCG without coevolution) we have two lines. The first one, depicted in black, is the theoretical line that differentiates which action is the \textit{risk-dominant} action (this is the line we called \textit{risk-dominant} transition line). The second one is the line that separates the region in which the system coordinates on action $A$ (green) from the region in which the system coordinates on action $B$ (yellow). This is the line we can refer to \textit{green-yellow} transition line. In this specific Figure, both lines are coincident and this is how we know that the system always coordinate on the \textit{risk-dominant} action. Our hypothesis is that coevolution displaces the second line towards the region in which action $B$ is the \textit{risk-dominant} action (the first line is theoretical, it does not change). A displacement of the second line towards the region in which action $B$ is the \textit{risk-dominant} while the first line remains in the same place would imply that there is an intermediate region in which the system coordinates on action $A$ for values of $(S,T)$ for which action $A$ is not the \textit{risk-dominant} action anymore. This is going to become clear in the next two sections.
\subsubsection*{Coordination on the Payoff-Dominant action}

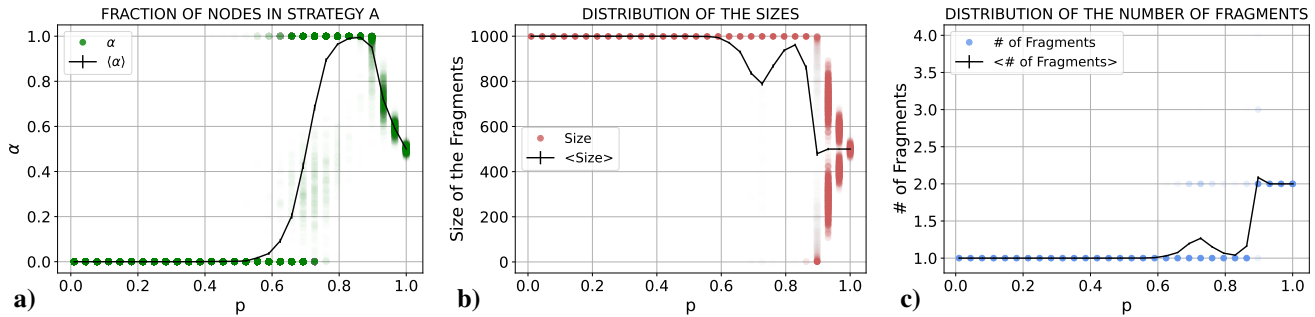
\begin{figure}[H]
    \centering
    \begin{minipage}[t]{0.33\linewidth}
        \begin{picture}(0.75\linewidth,0.75\linewidth)
            \put(0,0){\includegraphics[width=\linewidth]{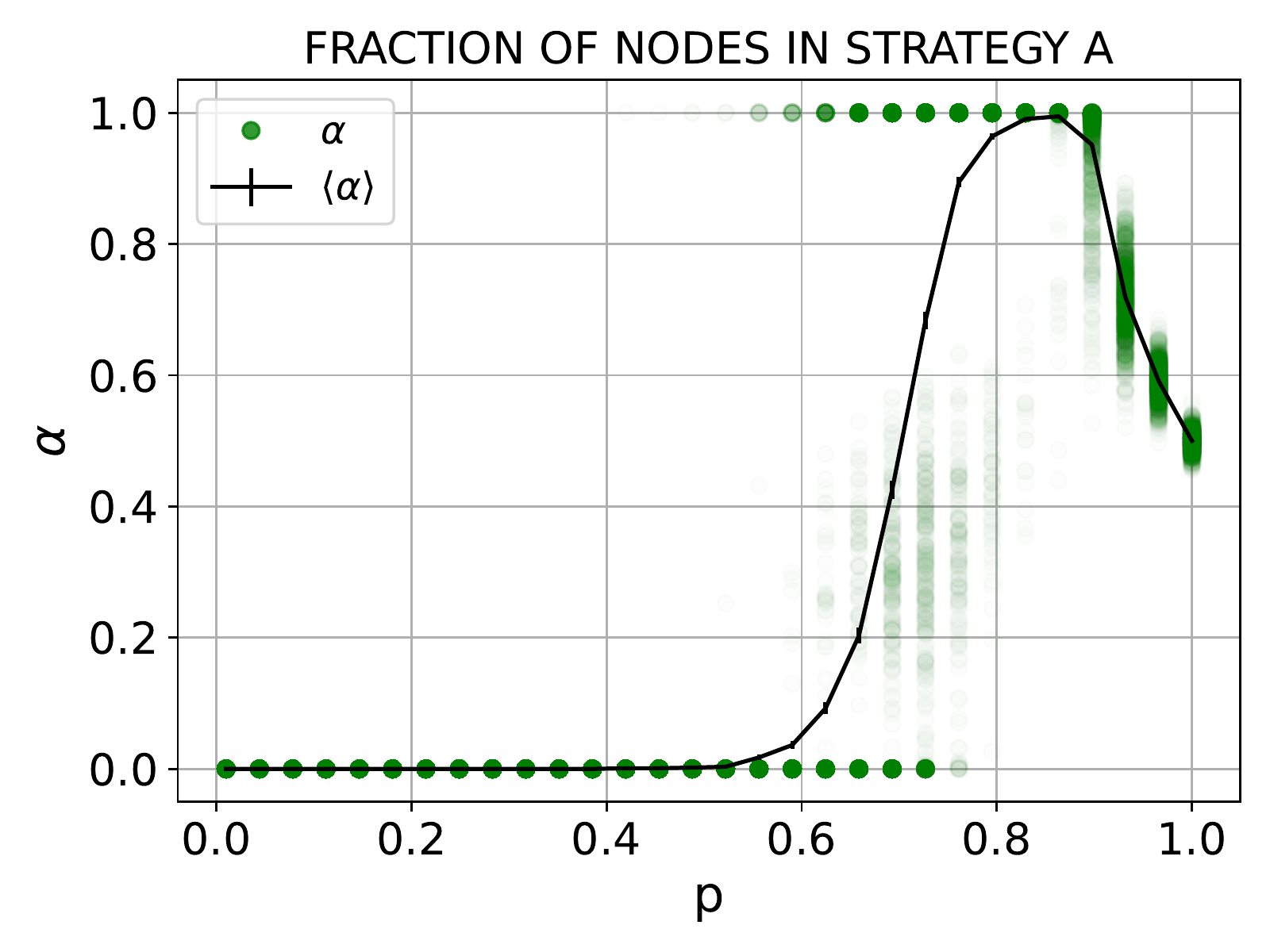}}
            \put(0.04\linewidth,0.04\linewidth){\textbf{a)}}
        \end{picture}
    \end{minipage}
    \begin{minipage}[t]{0.33\linewidth}
        \begin{picture}(0.75\linewidth,0.75\linewidth)
            \put(0,0){\includegraphics[width=\linewidth]{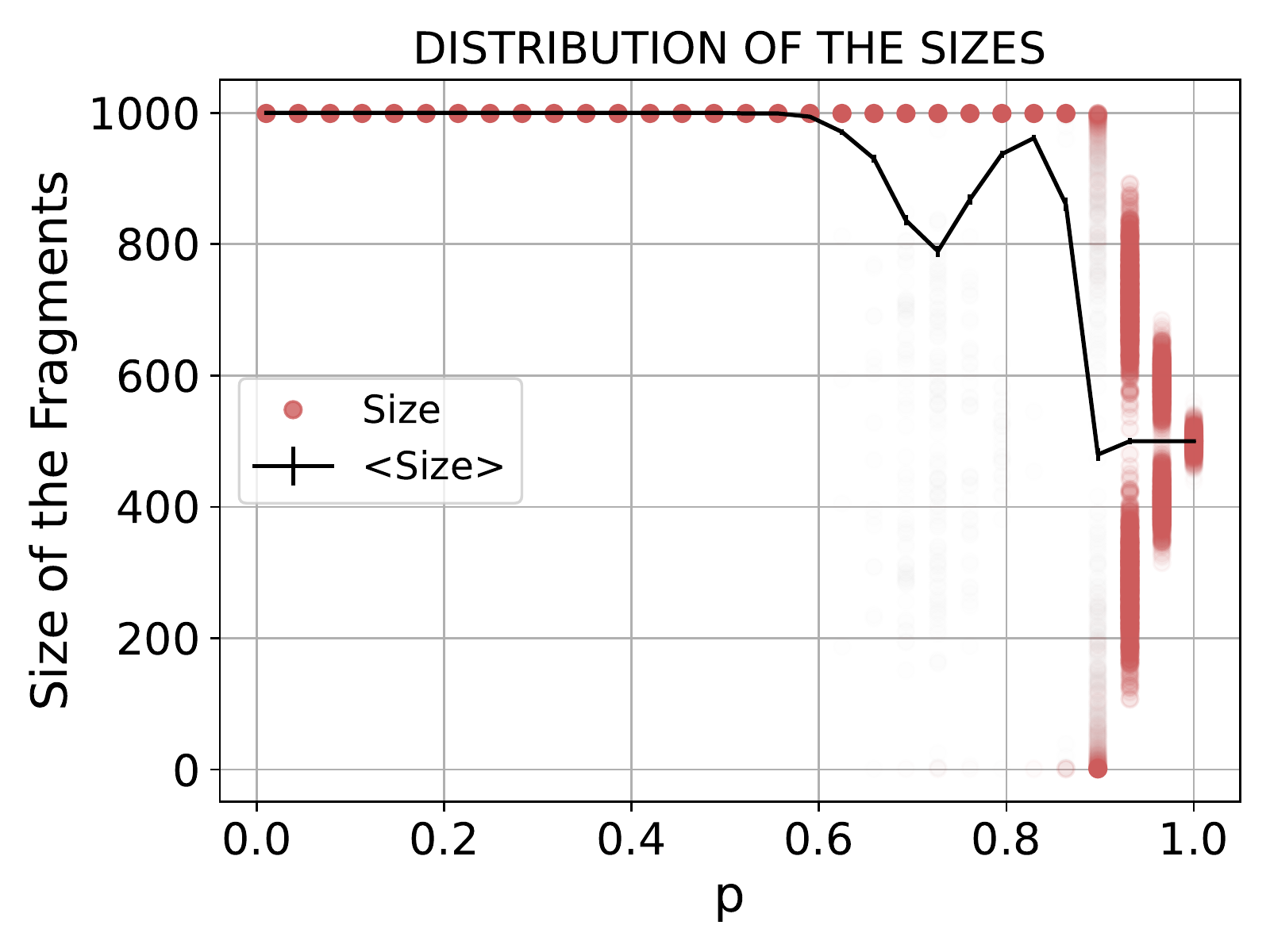}}
            \put(0.04\linewidth,0.04\linewidth){\textbf{b)}}
        \end{picture}
    \end{minipage}
    \begin{minipage}[t]{0.33\linewidth}
        \begin{picture}(0.75\linewidth,0.75\linewidth)
            \put(0,0){\includegraphics[width=\linewidth]{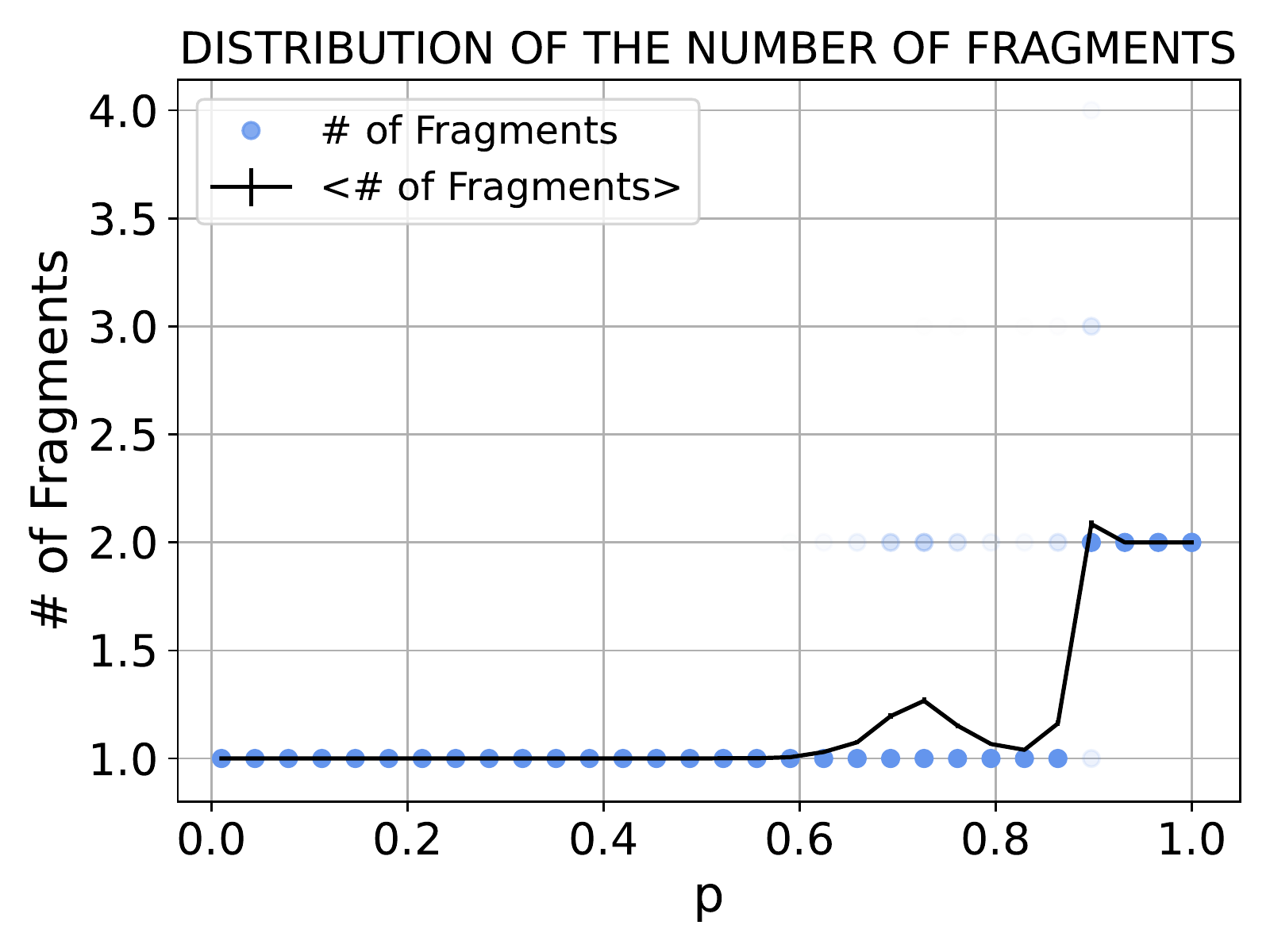}}
            \put(0.04\linewidth,0.04\linewidth){\textbf{c)}}
        \end{picture}
    \end{minipage}
    \caption{Results for a) $\alpha$, b) the size and c) the number of fragments (single realizations in green, red and blue, respectively, and average values in black) as a function of $p$ for a GCG with coevolution played in an E-R network with $\langle k \rangle=30$ using the UI update rule, with $(S,T)=(-1.35,0)$.}
    \label{GCG_C_payoff}
\end{figure}

Our hypothesis can be reformulated in the following way: for a certain region of the parameters space above and close to the \textit{risk-dominant} transition line (black line), where the \textit{payoff-dominant} action is different from the \textit{risk-dominant} action, coevolution allows the system to coordinate on the \textit{payoff-dominant} action. To test this hypothesis let us first compute $\alpha$, the size and the number of fragments for the fourth example pair $(S,T)$, which is precisely above and close to the black line (Fig. \ref{GCG_C_payoff}).

We can see that when $p$ is small the system coordinates on the \textit{risk-dominant} action, as before. Nonetheless, as we increase $p$, there is a transition to a regime in which the system fully coordinates on the \textit{payoff-dominant} action. At the transition, in some realizations the system fragments into two pieces, but in most of the cases it either coordinates on action $A$ or action $B$. After the transition the system always coordinates on the \textit{payoff-dominant} action. Finally, for high enough values of $p$ we have again the fragmentation transition we saw in the previous section. Indeed, these results show that coevolution displaces the transition line enabling the system to coordinate on the \textit{payoff-dominant} action even if this action is not the \textit{risk-dominant} action. This can be understood by realizing that a high enough value of $p$ enables agents using the \textit{payoff-dominant} action to connect with each other more often before they change to the \textit{risk-dominant} action due to the group pressure. Then, when they connect with each other they reinforce each other (since they receive a larger payoff than two agents using the \textit{risk-dominant} action connected between them). If the two actions are sufficiently similar (meaning that we are sufficiently close to the transition line of the $(S,T)$ space, precisely the case we are studying), this successful interaction compensates for their connection with agents using the \textit{risk-dominant} action and, thus, the \textit{payoff-dominant} action is able to take over. Anyway, this is just an example for a specific choice of $(S,T)$, The following step is naturally to visualize how the transition line actually changes, and this is the task we address in the following section.

\subsubsection*{Evolution of the Transition Line with the Rewiring}

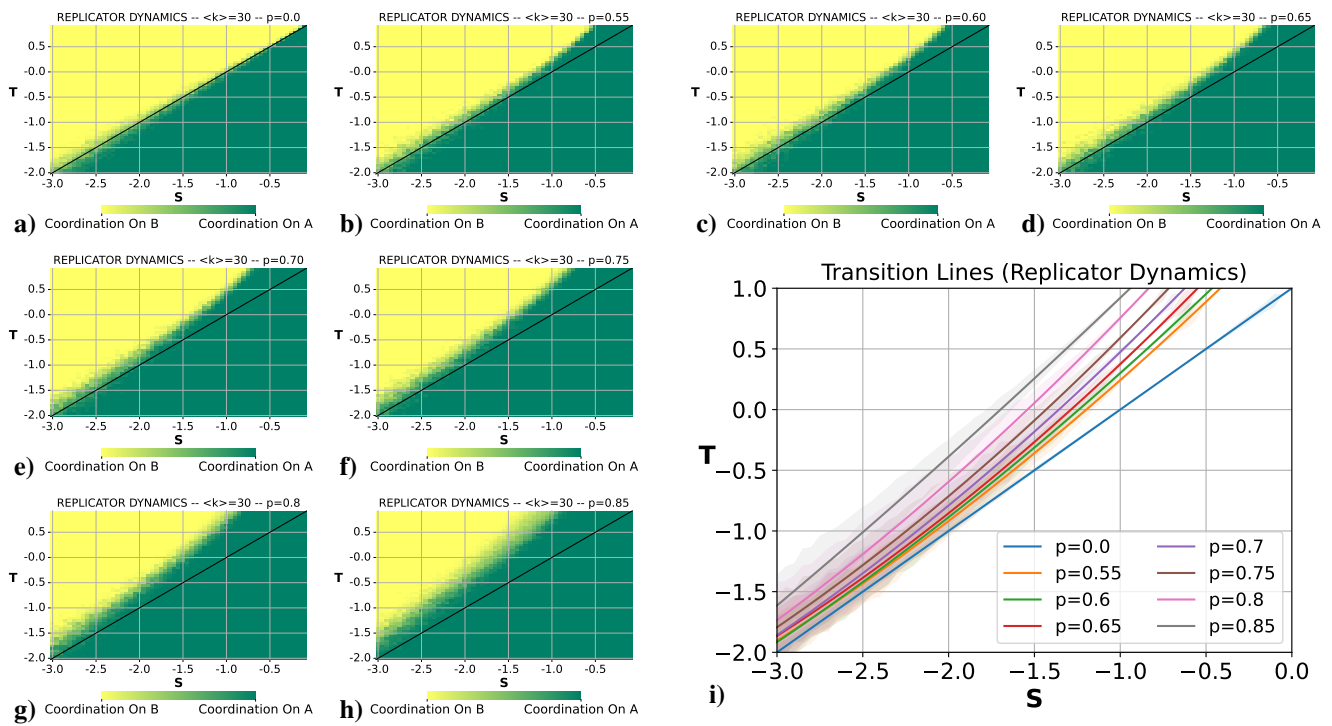
\begin{figure}[H]
\begin{multicols}{2}
    \begin{minipage}[t]{0.5\linewidth}
    \begin{picture}(0.75\linewidth,0.75\linewidth)
        \put(0,0){\includegraphics[width=\linewidth]{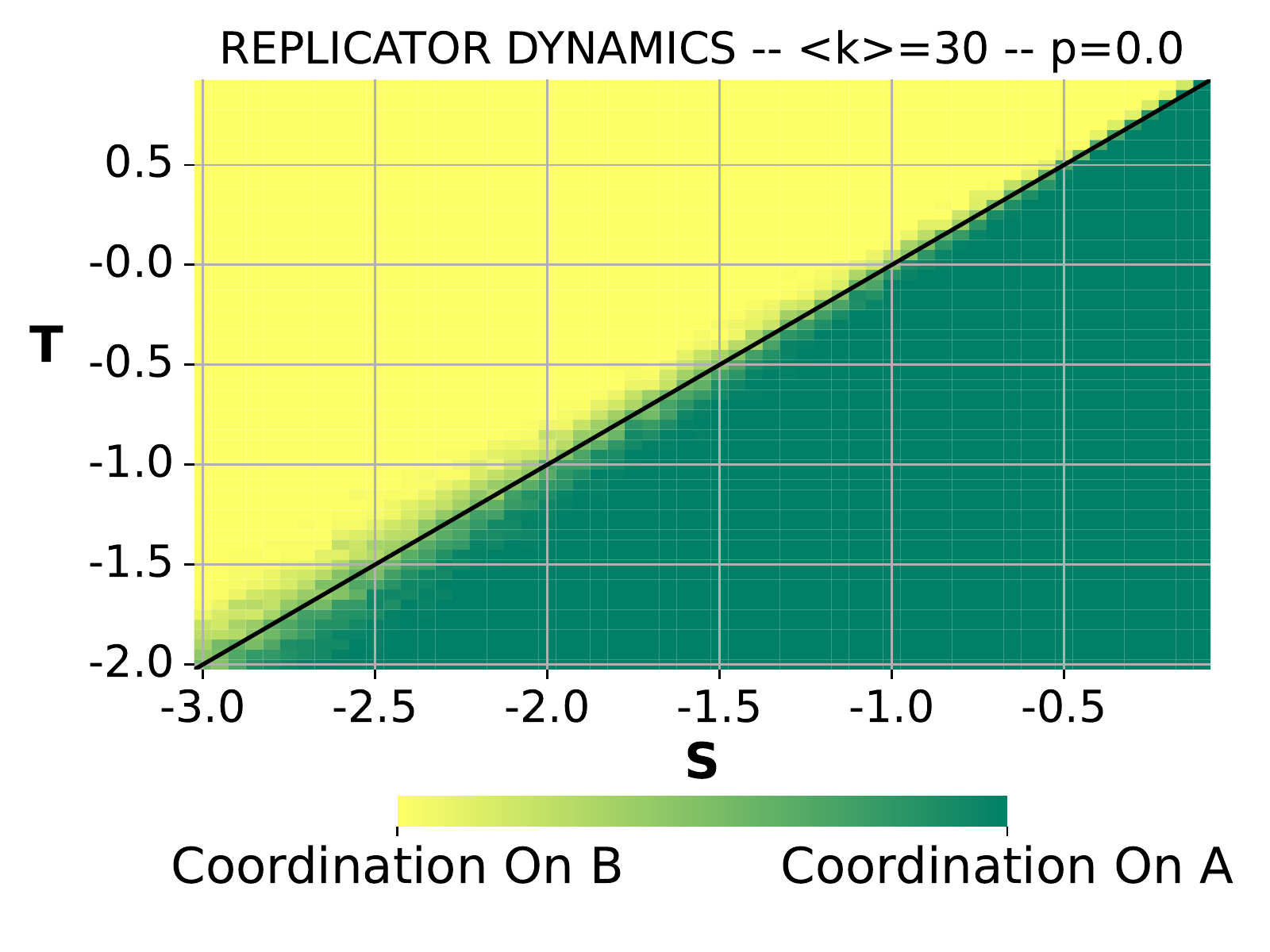}}
        \put(0.04\linewidth,0.04\linewidth){\textbf{a)}}
    \end{picture}
    \end{minipage}
    \begin{minipage}[t]{0.5\linewidth}
    \begin{picture}(0.75\linewidth,0.75\linewidth)
        \put(0,0){\includegraphics[width=\linewidth]{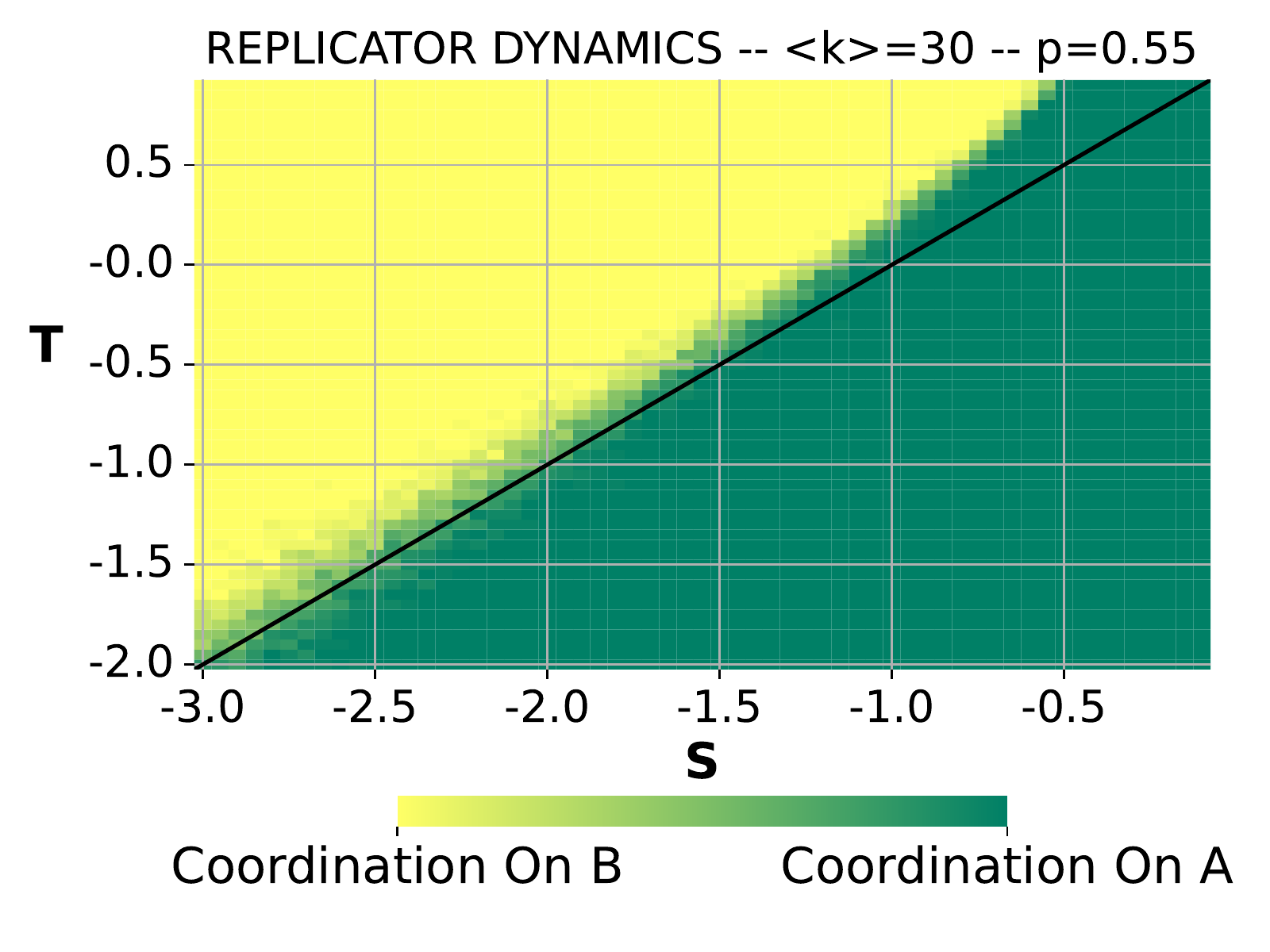}}
        \put(0.04\linewidth,0.04\linewidth){\textbf{b)}}
    \end{picture}
    \end{minipage}
    \begin{minipage}[t]{0.5\linewidth}
    \begin{picture}(0.75\linewidth,0.75\linewidth)
        \put(0,0){\includegraphics[width=\linewidth]{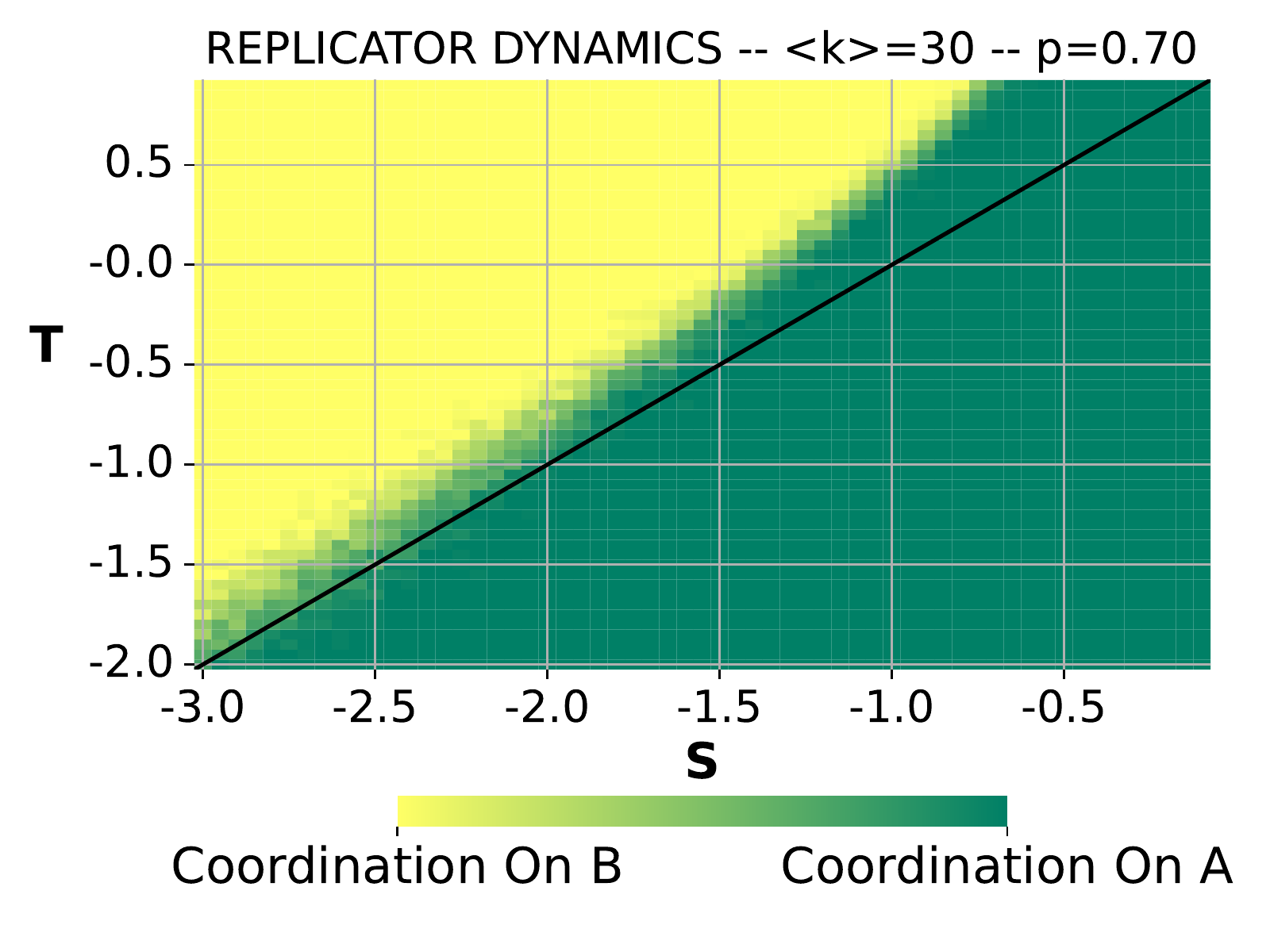}}
        \put(0.04\linewidth,0.04\linewidth){\textbf{e)}}
    \end{picture}
    \end{minipage}
    \begin{minipage}[t]{0.5\linewidth}
    \begin{picture}(0.75\linewidth,0.75\linewidth)
        \put(0,0){\includegraphics[width=\linewidth]{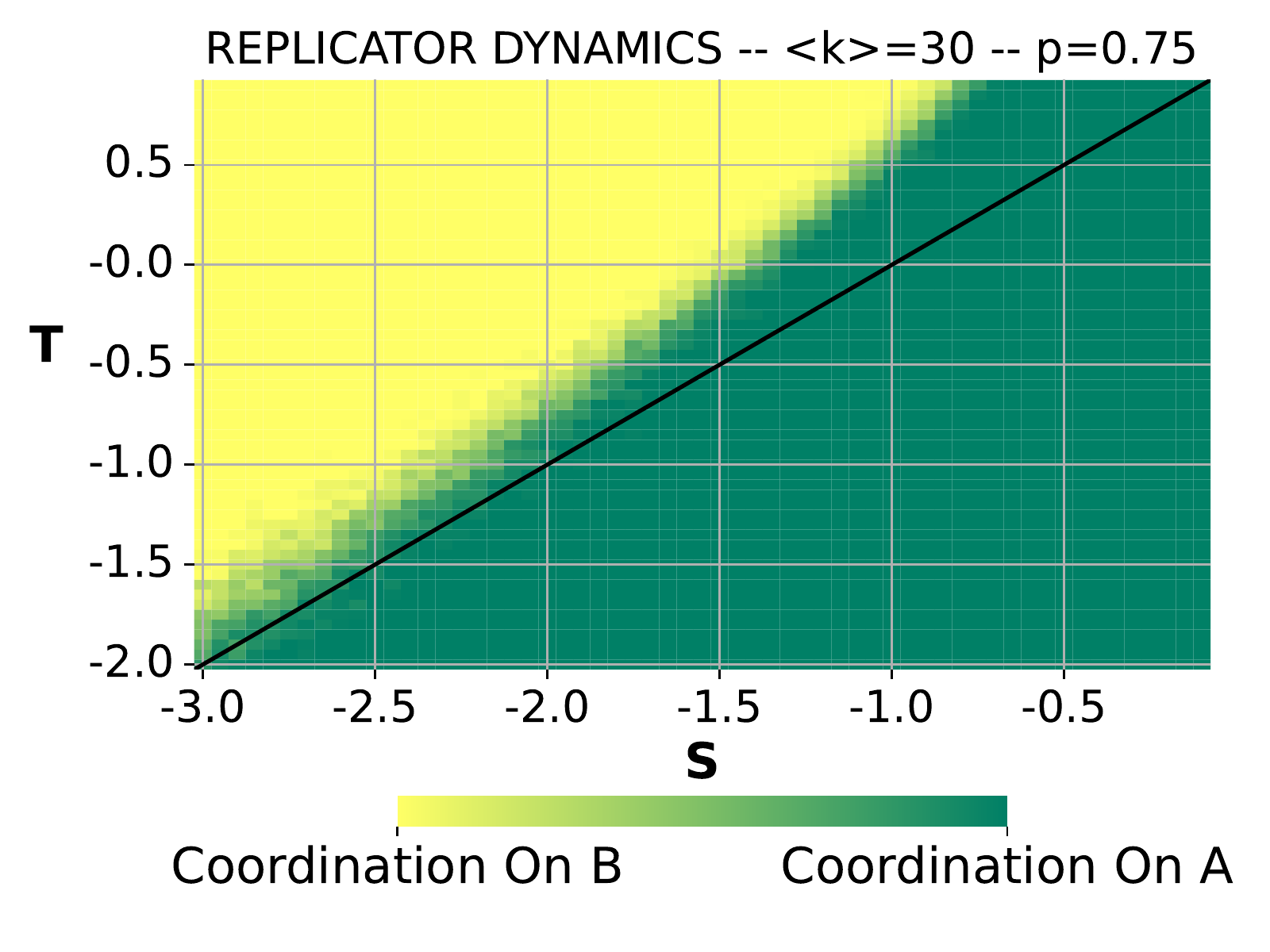}}
        \put(0.04\linewidth,0.04\linewidth){\textbf{f)}}
    \end{picture}
    \end{minipage}
    \begin{minipage}[t]{0.5\linewidth}
    \begin{picture}(0.75\linewidth,0.75\linewidth)
        \put(0,0){\includegraphics[width=\linewidth]{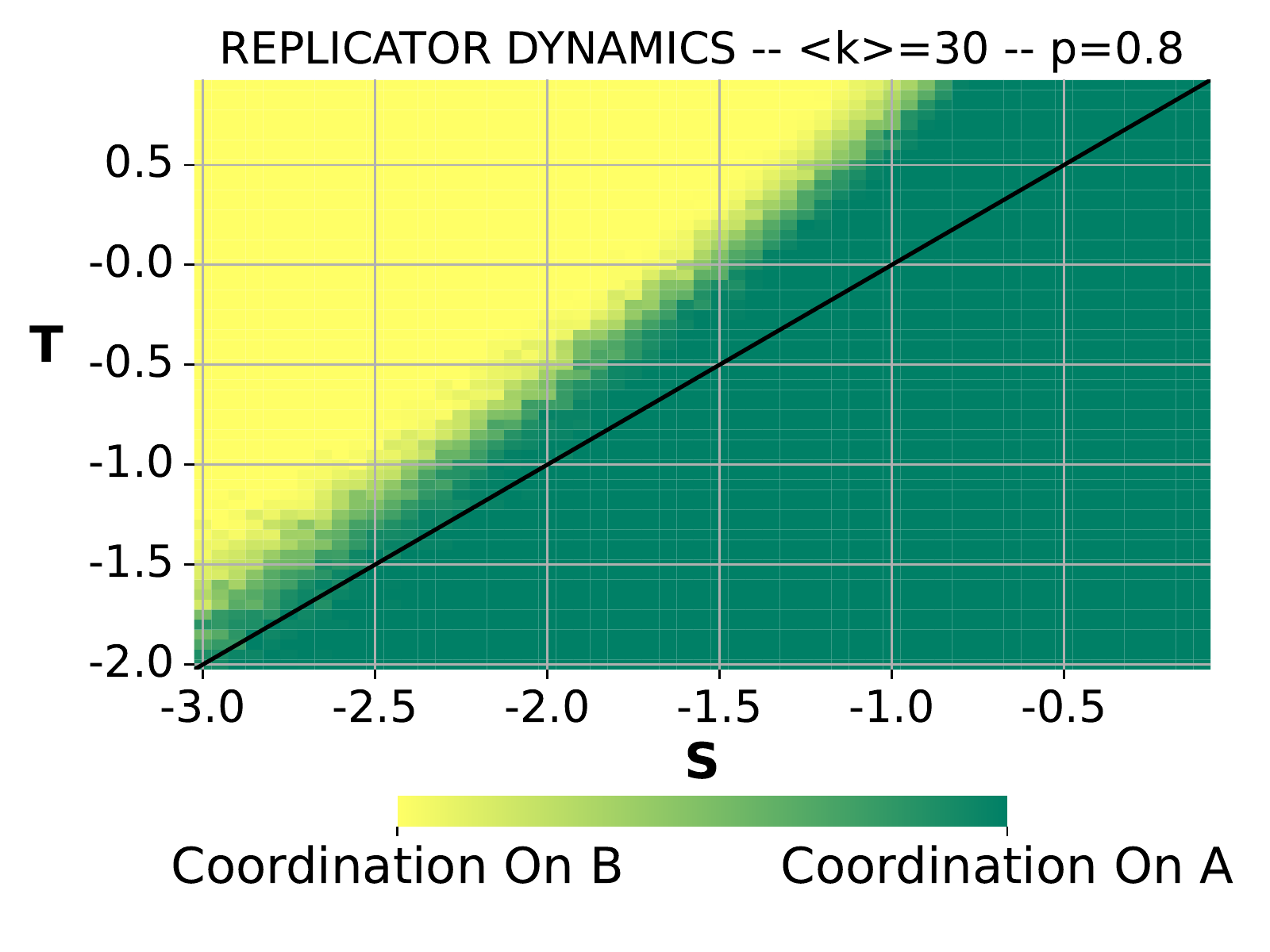}}
        \put(0.04\linewidth,0.04\linewidth){\textbf{g)}}
    \end{picture}
    \end{minipage}
    \begin{minipage}[b]{0.5\linewidth}
    \begin{picture}(0.75\linewidth,0.75\linewidth)
        \put(0,0){\includegraphics[width=\linewidth]{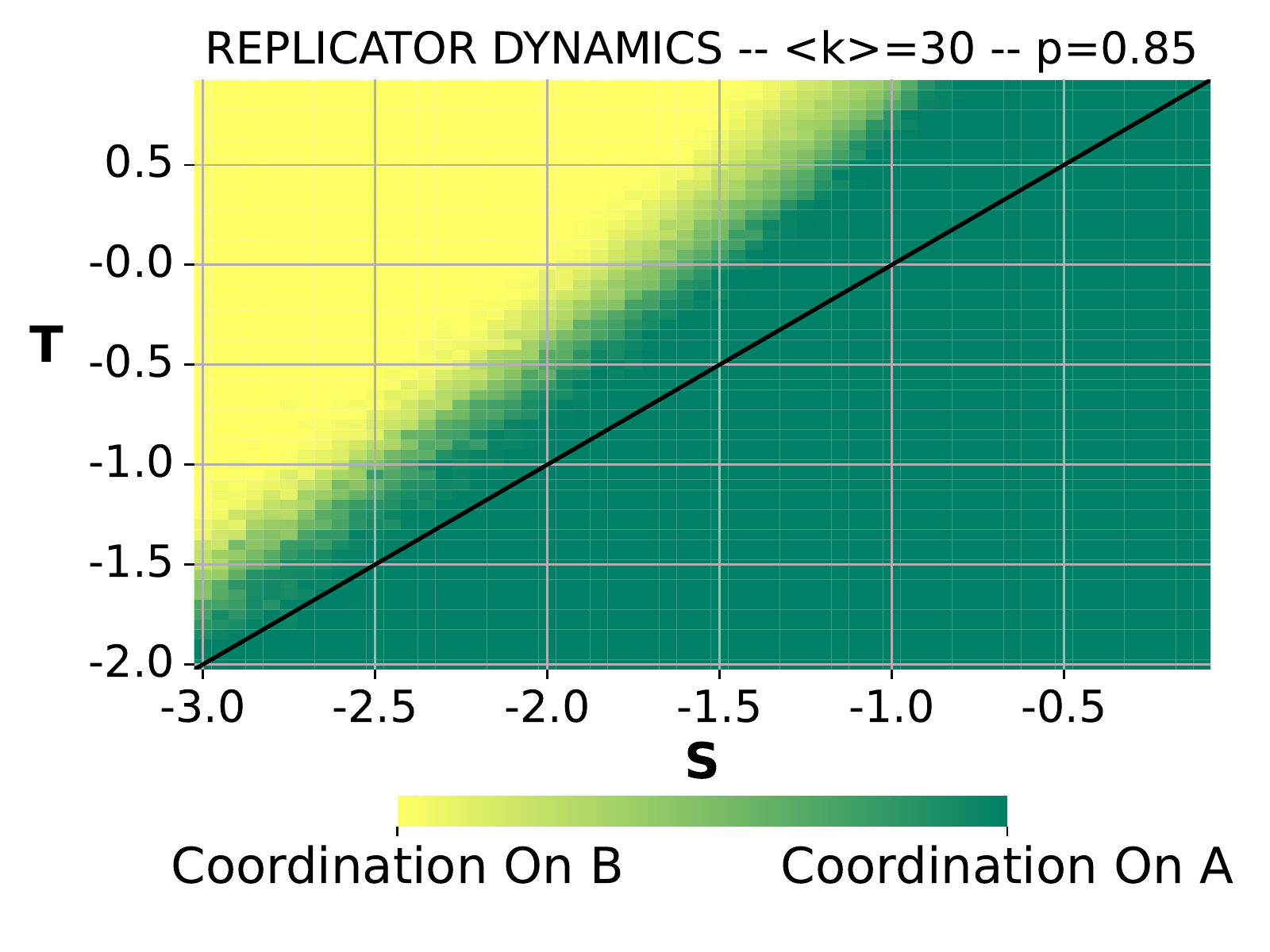}}
        \put(0.04\linewidth,0.04\linewidth){\textbf{h)}}
    \end{picture}
    \end{minipage}
    \vfill\columnbreak
    \begin{minipage}[t]{0.5\linewidth}
    \begin{picture}(0.75\linewidth,0.75\linewidth)
        \put(0,0){\includegraphics[width=\linewidth]{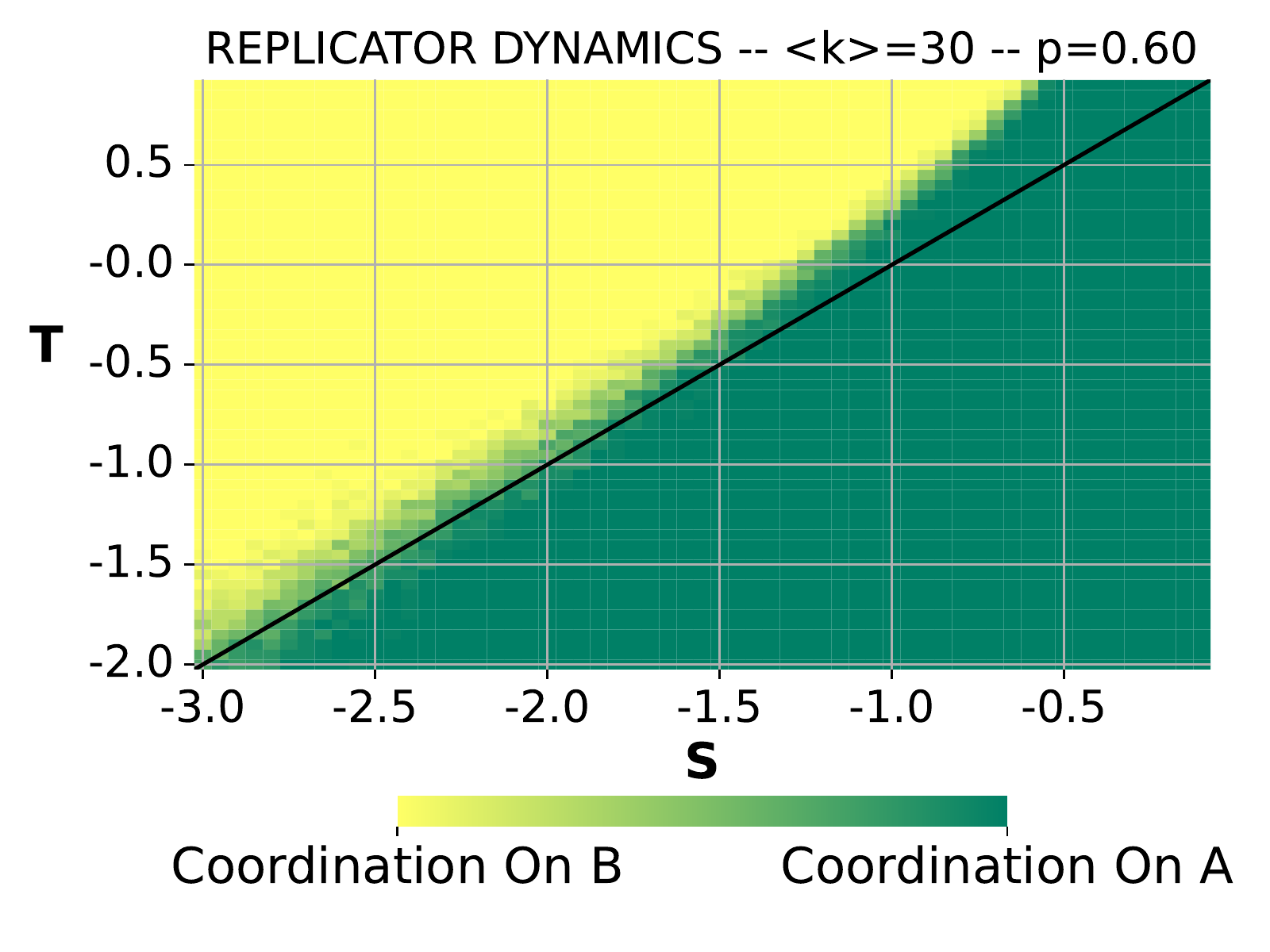}}
        \put(0.04\linewidth,0.04\linewidth){\textbf{c)}}
    \end{picture}
    \end{minipage}
    \begin{minipage}[t]{0.5\linewidth}
    \begin{picture}(0.75\linewidth,0.75\linewidth)
        \put(0,0){\includegraphics[width=\linewidth]{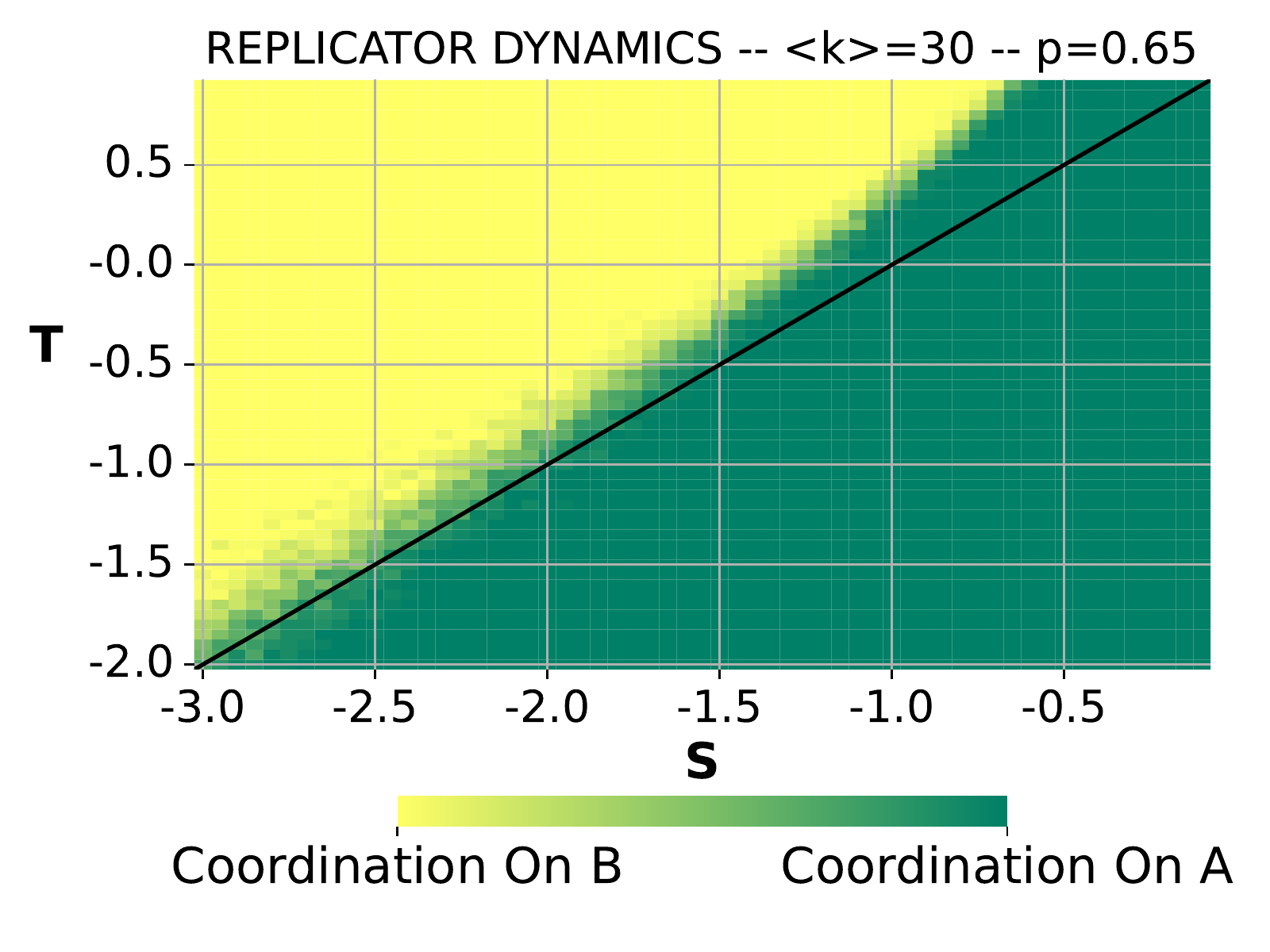}}
        \put(0.04\linewidth,0.04\linewidth){\textbf{d)}}
    \end{picture}
    \end{minipage}
    \begin{minipage}[t]{\linewidth}
    \begin{picture}(0.75\linewidth,0.75\linewidth)
        \put(0,0){\includegraphics[width=\linewidth]{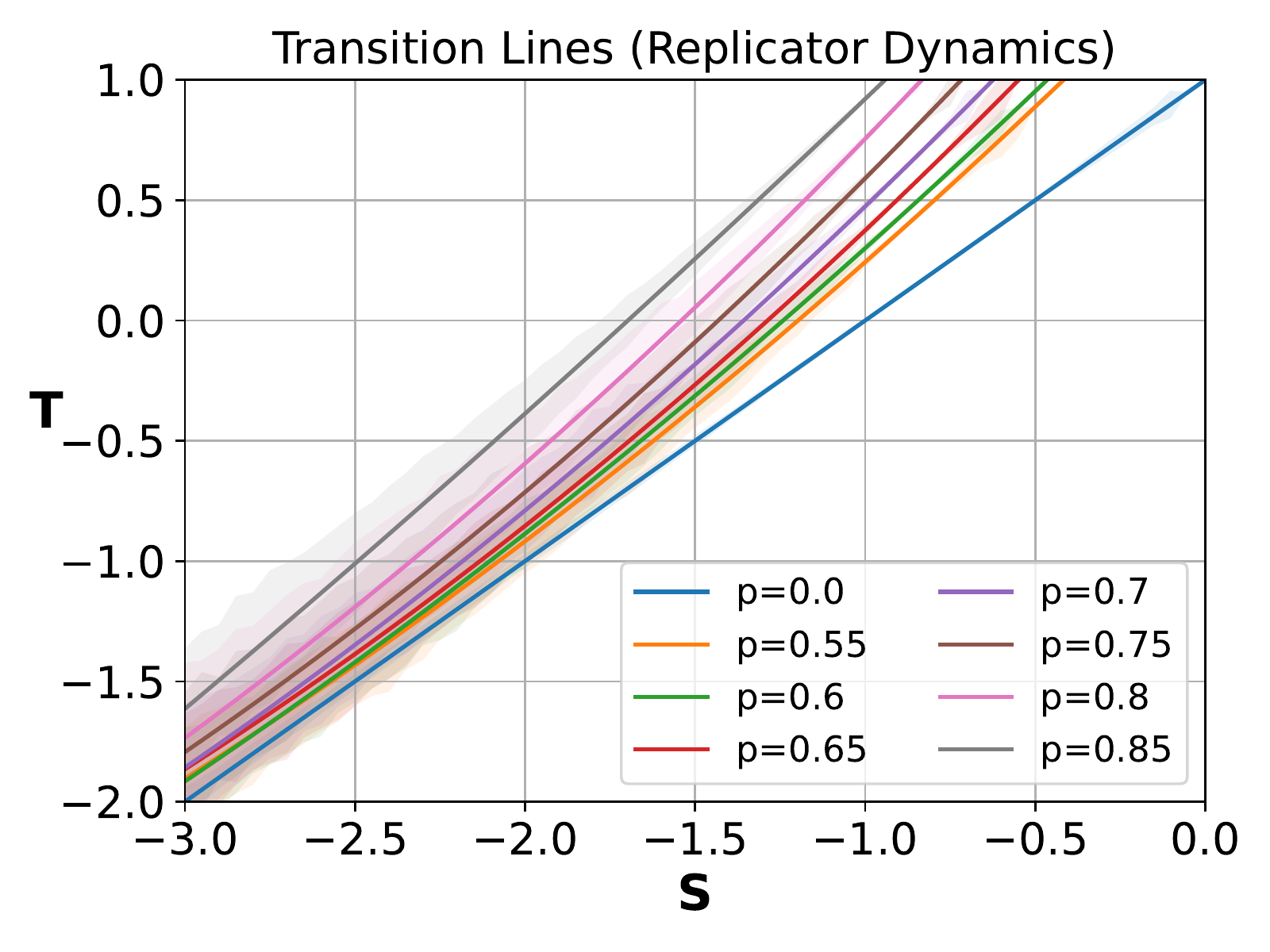}}
        \put(0.04\linewidth,0.04\linewidth){\textbf{i)}}
    \end{picture}
    \end{minipage}
\end{multicols}  
\caption{\textbf{a)-h) Panels:} $(S,T)$ diagrams showing the average value of $\alpha$ for different choices of $p$. \textbf{i) Panel:} Summary of the \textit{green-yellow} transition lines obtained for different choices of $p$.}
\label{GCG_C_evolution1}
\end{figure}

In order to complete this analysis, we select a group of values of $p$ and study how the transition line actually changes with $p$. The idea is to select values of $p$ low enough to avoid the regime in which the system fragments (we are only interested in the system coordinating on a single connected component) but high enough for the changes in the transition line to be noticeable. We collect in Fig. \ref{GCG_C_evolution1} the results obtained. In this Figure, we depict analogous diagrams to the Figure included in the Supplementary Material depicting the equilibrium selection within the parameter space, now for the case with coevolution for different values of p (panels a)-h)), and we include a plot summarizing all the \textit{green-yellow} transition lines obtained for different values of $p$ (panel i)). It can be clearly seen that the hypothesis was correct, and that the transition line displaces towards the region in which action $B$ is the \textit{risk-dominant} action, leaving an intermediate region in the parameters space (between the \textit{risk-dominant} transition line and the \textit{green-yellow} transition line) in which the system is able to fully coordinate on the \textit{payoff-dominant} action even if it is not the \textit{risk-dominant} action.

\begin{figure}[H]
    \centering
    \begin{minipage}[t]{0.33\linewidth}
        \includegraphics[width=\linewidth]{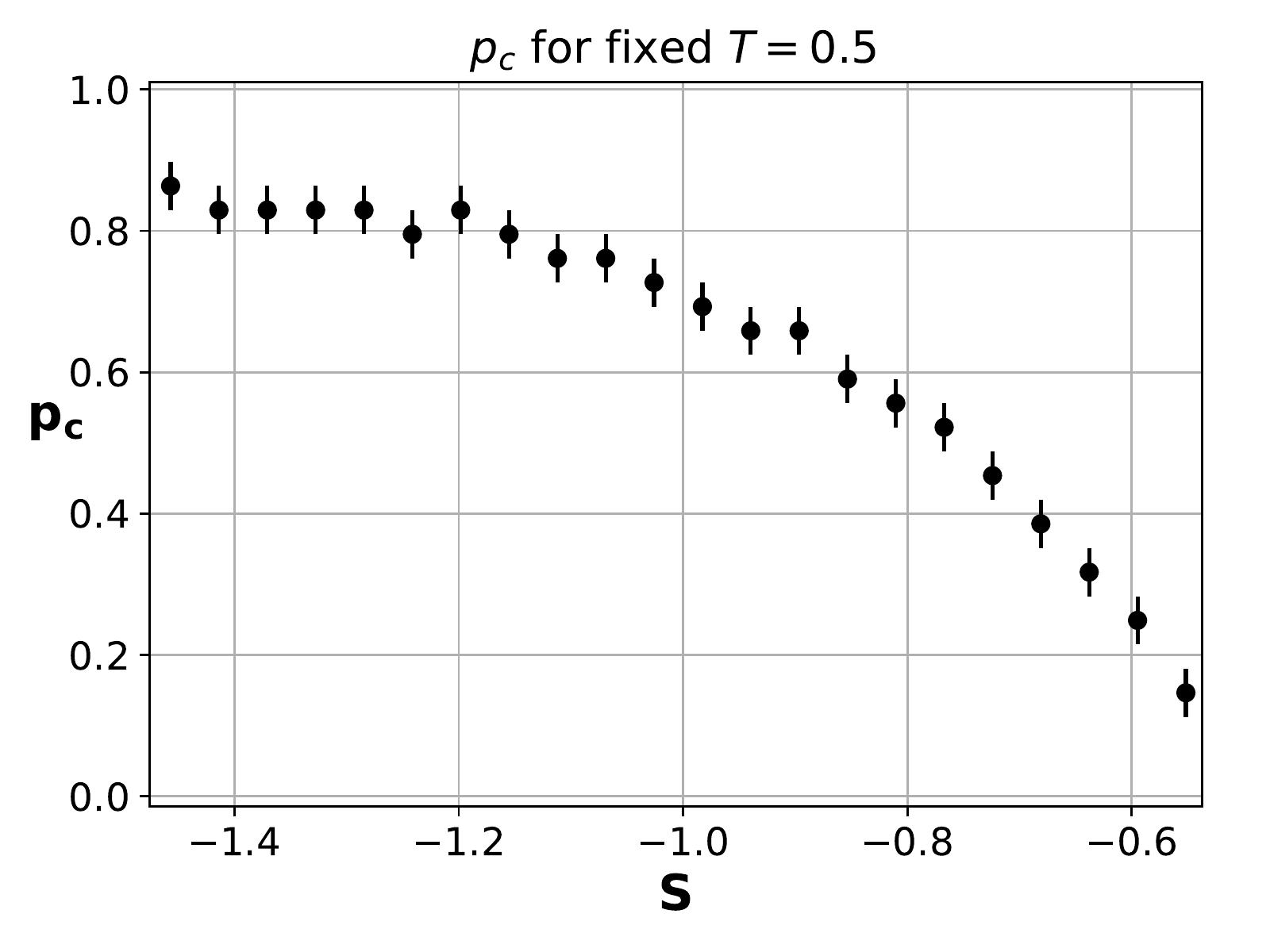}
    \end{minipage}
    \caption{Value $p_c(S)$ for which we have the transition from the \textit{risk-dominant} action to the \textit{payoff-dominant} one.}
    \label{GCG_C_evolution2}
\end{figure}

Finally, to complete the picture we conduct a final analysis in which we fix $T=0.5$ and, for values of $S$ inside the range $(-1.5,-0.5)$ we compute the critical value $p_c$ for which we have the transition from coordination on the \textit{risk-dominant} action to coordination on the \textit{payoff-dominant} one. This value $p_c$ is computed as the peak in the variance of the distribution of $\alpha$. We restrict ourselves to this small range of $S$ due to the fact that for $S>-0.5$ action $A$ becomes \textit{risk-dominant}. This implies that for $S>-0.5$ $A$ is both \textit{risk} and \textit{payoff-dominant} and hence there is no transition from one to the other. On the other hand, for $S<-1.5$ we directly have the fragmentation transition when we increase $p$, since the system is not able to coordinate on the \textit{payoff-dominant} action. We present in Fig. \ref{GCG_C_evolution2} the result obtained. This plot shows how the closer we get to the \textit{risk-dominant} transition line, the smaller the value of the rewiring needed to enable the system to coordinate on the \textit{payoff-dominant} action. This result is expected since the closer we are to the transition line the more similar both actions are, and hence the "easier" it is to coordinate on the \textit{payoff-dominant} action.
All the analysis performed for the RD case was conducted as well for the UI case. Analogous results can be found on the Supplementary Material (Sections S4 and S5). For the UI case we observe the fragmentation transition as in the RD case, and this fragmentation transition is again analogous to the one present for a PCG (fragmentation from one single connected component to two connected components with an intermediate region in which the system fragments in multiple components). It can be seen as well that the transition line is displaced towards the region in which action $B$ is the \textit{risk-dominant} action. Nonetheless, this case is not as interesting as the RD case due to the fact that this region in which the system is able to coordinate on the \textit{payoff-dominant} action already existed in the absence of coevolution. With coevolution this region is simply broadened. As a collective conclusion for both update rules, coevolution enlarges the region in which the system is able to fully coordinate on the \textit{payoff-dominant} action with respect to the case without coevolution.

\section*{Discussion}

In this paper we have assessed the role of the network coevolution within Coordination Games. We have focused on values of the mean degree $\langle k \rangle$ of a random network for which the system is, in principle, able to reach full coordination in the absence of coevolution. First, we have explored the case of Coordination Games with equivalent actions, Pure Coordination Games. We have found that, taking the network plasticity $p$ as a control parameter, there is a fragmentation transition both for the RD and the UI update rules. For the RD rule, there is a \textit{simple} transition from coordination on a single connected component in any action with equal probability to a regime in which the system fragments in two connected components, each of them coordinated on one of the actions. Close to the transition one of the components is larger than the other, and they become of a similar size as we approach $p=1$. This transition is similar to the one observed, for instance, in the Coevolving Voter Model\cite{CoevolvingVoterModel}, and for the Prisoner's Dilemma with coevolution\cite{timescale6}. 
Conversely, for the UI update rule the nature of the transition is quite different. For low $p$ we observe that the cases in which the system was able to fully coordinate in the absence of coevolution continue doing so, whereas most of the cases that ended up in a frozen state without coevolution now end up in a fragmented state. For $p$ close to $1$ we have a regime in which the system fragments in two components as well. Nonetheless, for intermediate values of $p$ we now have fragmentation in multiple components: first, only small components plus a large connected component and then two big components plus small components. This kind of shattered fragmentation has been observed as well in different versions of a  Coevolving Voter Model\cite{PaperMarina}${}^{,}$\cite{PaperMarina2}.  

We have also investigated Coordination Games with non equivalent actions, General Coordination Games.  We have found that the system undergoes a fragmentation transition similar to the one observed in Pure Coordination Games, both for RD and UI. This allows a part of the system to coordinate on the \textit{payoff-dominant} action even if this action is not the \textit{risk-dominant} action, contrary to what we observe without coevolution. Perhaps more importantly, we find an intermediate regime of values of the rewiring probability $p$, before the fragmentation transition, for which the system is able to fully coordinate on the \textit{payoff-dominant} action in a single network component. Interestingly, these results are compatible with previous ones for different coevolution models that also obtained that coevolution can enhance coordination on the \textit{payoff-dominant} action\cite{CoevolutionCoordination2}${}^{,}$\cite{CoevolutionCoordination3}${}^{,}$\cite{Pestelacci}. It is also worth mentioning that in a previous study \cite{efficientconventions} of a general Coordination Game with the UI rule, it was shown that if interactions are sporadic, the system tends to coordinate on the \textit{payoff-dominant} action more than if interactions are more frequent. This is, again, compatible with our results: a high $p$ plasticity makes interactions within the Coordination Game more sporadic, and it is precisely for higher values of $p$ for which we observe coordination on the \textit{payoff-dominant} action.

It is important to outline the limits of our results. First of all, we have explored only ER networks. Nonetheless, other network structures could introduce new unobserved effects. Also, these networks are generated with the same number of nodes in a narrow range of link density, so we may be overlooking finite-size effects. Finally, our results are based on computer simulations, with the inherent limitations on the potential conclusions that can be drawn compared to work substantiated by analytical proofs. Anyway, we believe studies like ours introduce new phenomena and foster further analytical and more rigorous work about these new phenomena, as well as further analysis using different network structures and sizes.

In addition, future efforts along these lines could focus both on experimental research to identify, for instance, which update rules better reproduce real human behavior, and on the implementation of rules that we already know mimic human decision-making processes. For instance, in works like \cite{LearningProcess} they uncovered a learning process in humans playing a spatial Prisoner's Dilemma, and in \cite{PrincipalAnxo} they developed a model which implements this learning process in an update rule (Reinforcement Learning), successfully reproducing the results of these experimental studies. Building upon previous experimental results on Coordination Games\cite{ExperimentalCoordination}${}^{,}$\cite{anxoexperimentcoord}, it would be interesting to develop a similar learning rule both for the action update in a Coordination Game, and for the process of coevolution (agents learning both after playing the game and after changing connections). In this respect, a starting point could be the work of Lozano et al.\cite{lozano} in which they introduced behavioral rules for the link making and breaking decisions that reproduced well the experimental results. All in all, we can conclude that coevolution within the context of Coordination Games is a powerful mechanism with the potential to explain the emergence of coordinated groups within a population. Furthermore, it is able to enhance coordination on riskier but socially and individually profitable actions, even in systems for which this is not possible without coevolution.

\section*{Methods}\label{methods}

In general, for the different analysis conducted in this paper, we follow a similar methodology. First, we define the coevolutionary rule, that describes what happens at each time step in the system, and then we make use of this coevolutionary rule to describe how we perform the analysis.

\subsection*{Coevolution Rule within a Coordination Game} \label{implementation}

The two rules chosen for this paper are Replicator Dynamics and Unconditional Imitation. As we mentioned in the introduction, the purpose of this paper was to study the consequences of coevolution of the network on coordination and equilibrium selection, building upon the results for static networks. Specifically, our work considers as a point of departure the investigation conducted by T. Raducha and M. San Miguel (2022). Thus, for comparative purposes, we selected the same rules as them. In summary, these two rules work as follows:\\
\textbf{Replicator Dynamics update rule}\\
Let us consider that our agent $i$ is using action $s_i$, and its randomly chosen neighbor $j$ uses action $s_j$. If this neighbor is using the same action than our agent, nothing happens. Otherwise, if it is using another action, our agent compares their payoffs. If the neighbor's payoff is larger, then our agent copies its neighbor's action with a probability given by:
\begin{equation}
    P(s_i\rightarrow s_j)=\frac{\pi_j-\pi_i}{\Phi}
\end{equation}
Where $\pi_{i,j}$ is the payoff obtained by agents $i, j$ after playing with all of their neighbors, and $\Phi$ is the maximum payoff difference there can be, to ensure that the probability is in the range $[0,1]$. This rule states that the probability of a action reproducing  is larger if the payoff difference between both agents is larger.\\
\textbf{Unconditional Imitation update rule}\\
In this rule, our agent $i$ compares its payoff with the payoff of all of its neighbors. Then, $i$ chooses the most successful neighbor (i. e. the one with the largest payoff) and copies its action.\\
With these two rules in mind, we construct the coevolutionary rule adding the rewiring probability. Specifically:\\
At each time step, we select randomly an agent $i$.
\begin{itemize}
    \item We select randomly one of its neighbors $j$. If both agents use a different action, with a certain probability $p$ (rewiring probability) agent $i$ cuts the link with agent $j$ and creates another connection with an agent randomly chosen from the rest of the population, irrespective of the action the other agent is currently using.
    \item If the agent does not rewire, we apply the RD or UI update rule, depending on the case we are at.
    \item Everyone updates its aggregated payoff.
\end{itemize}
Notice that both implementations are slightly different in the way we select which agent to imitate and which agent to cut connection with. Specifically, the difference is that, after selecting the agent $i$:
\begin{itemize}
    \item Replicator Dynamics: we select \textit{randomly} $j$, one of its neighbors, and, whatever happens after, either rewiring or imitation, is referred to this neighbor. 
    \item Unconditional Imitation: we select \textit{randomly} $j$, one of its neighbors, and we evaluate if there is rewiring. If not, our agent imitates the best performing neighbor, even if this neighbor is not $j$
\end{itemize}
 We outline this difference because there is another way to define the UI rule with rewiring. We could, first, select the neighbor $j$ with the largest payoff and, then, if this neighbor used a different strategy, apply rewiring with a certain probability, and imitate its strategy if there is no rewiring. But, with this rule, agent $i$ is never able to sever a connection with a neighbor using a different strategy if this neighbor does not have the largest payoff. Thus, the system would never converge to state with zero density of active links (see discussion below). Instead, we define the rule in a way that enables agent $i$ to sever any connection or imitate the best performing neighbor.
\subsection*{Simulation details }
First we generate a network with an Erdös-Rényi structure\cite{Estrada1}. This network has $N=1000$ nodes, and it is generated using the link formation probability $p=\langle k \rangle/(N-1)$, where $\langle k \rangle$ is the mean degree of the network considered. Second, we distribute randomly among the nodes actions $A$ and $B$ in equal proportion. Before starting to evolve in time, all agents play the game with all of the neighbors they are connected to, and they accumulate all the payoffs from those pairwise games. Then, at each time step we select one agent randomly and we make its action evolve according to the coevolutionary rule specified above. If this evolution leads to a change in its action, this agent and its neighbors update their payoffs due to the change on the action (they forget their previous payoff and they play again accumulating the total payoff from each pairwise interaction). 

We define the density of active links as the number of links connecting agents that use different actions divided by the total number of links in the network, and we call it $\rho$. This variable determines whether the system has reached a frozen state. Specifically when this density of links connecting agents in different states is zero, it means that all agents are connected to other agents using their strategy, i. e. they are coordinated with their neighbors, and the system freezes since they have no longer incentive to change state according to the model. In other words, if this density of active links is zero it means that each connected component of the network is coordinated on one of the actions and, therefore, none of the agents can update their action. All in all, we let the system evolve following these rules until it reaches a frozen state.  We can have two types of frozen states: on the one hand, the system could reach a state in which there is a single connected component in the network. This outcome corresponds to the situation in which the whole system coordinates on a single action. If this is the case, we say that the system has reached \textit{full coordination}. The other possibility is to have multiple connected components, i. e., the system freezes in a state where actions $A$ and $B$ coexist, because there are different connected components in the network, each of them fully coordinated on one of the actions.

The explained process is what we call a single \textit{realization}: we set up the system and let it evolve until it reaches a frozen state. Then, in order to make measurements of the different statistical properties of the system, we repeat this simulation several times, measure the relevant magnitudes and average the results over realizations. The exact number of realizations will be specified in each case. Now, we define the most important magnitudes that we will measure. On the one hand, we want to know if the system reaches full coordination or not. To assess it, we define $\alpha$ as the proportion of nodes that use action $A$. It is clear then that $\alpha \in [0,1]$, where $0$ corresponds to full coordination on $B$ and $1$ corresponds to full coordination on $A$. On the other hand, we measure the number of connected components we have on the network, and the sizes of these components.

\section*{Acknowledgements}

MSM acknowledges support from projects PACSS (reference RTI2018-093732-B-C22), APASOS (reference PID2021-122256NB-C21)  and the Maria de Maeztu program for Units of Excellence in R\&D (MDM-2017-0711) financed by MCIN/AEI/10.13039/501100011033 / FEDER, UE. AS acknowledges support from project BASIC (PGC2018-098186-B-I00) funded by MCIN/AEI/10.13039/501100011033 and by “ERDF A way of making Europe”.

\section*{Author contributions statement}

M.S.M. and A.S. conceived the research, M.A.G.C. wrote the computer code, run the simulations and analyzed the data, and all authors discussed and interpreted the results and wrote the manuscript.

\section*{Data availability statement}
Data and computer codes are available from M.A.G.C. upon reasonable request.
\section*{Ethics declarations}

\subsection*{Competing interests}

The authors declare no competing interests.

\end{document}